\documentclass[twocolumn]{aastex63}
\usepackage{epstopdf}
\usepackage{amsmath}
\usepackage{amssymb}
\usepackage{natbib}
\usepackage{bbold}
\usepackage{listings}
\usepackage{dsfont}
\usepackage{xcolor}

\newcommand{\matr}[1]{\mathsf{#1}}
\newcommand{\vect}[1]{\mathbf{#1}}

\begin{document}

\title{Direction Dependent Corrections in Polarimetric Radio Imaging III: A-to-Z Solver - Modeling the full Jones antenna aperture illumination pattern}
\author{Srikrishna Sekhar} \affiliation {Inter-University Institute for Data Intensive Astronomy, and Department of Astronomy, University of Cape Town, Cape Town, South Africa}
 \affiliation{National Radio Astronomy Observatory, 1003 Lopezville Road, Socorro, NM 87801, USA}  \affiliation{Inter-University Institute for Data Intensive Astronomy, and Department of Astronomy, University of the Western Cape, Bellville, South Africa}

\author{Preshanth Jagannathan}\affiliation{National Radio Astronomy Observatory, 1003 Lopezville Road, Socorro, NM 87801, USA}
\author{Brian Kirk}\affiliation{National Radio Astronomy Observatory, 1003 Lopezville Road, Socorro, NM 87801, USA}
\author{Sanjay Bhatnagar}\affiliation{National Radio Astronomy Observatory, 1003 Lopezville Road, Socorro, NM 87801, USA}
\author{Russ Taylor} \affiliation {Inter-University Institute for Data Intensive Astronomy, and Department of Astronomy, University of Cape Town, Cape Town, South Africa}
\affiliation{Inter-University Institute for Data Intensive Astronomy, and Department of Astronomy, University of the Western Cape, Bellville, South Africa}

\begin{abstract}

In this third paper of a series describing direction dependent corrections for
polarimetric radio imaging, we present the the A-to-Z solver methodology to
model the full Jones antenna aperture illumination pattern (AIP) using Zernike
polynomials. In order to achieve accurate, thermal noise limited imaging with
modern radio interferometers, it is necessary to correct for the instrumental
effects of the antenna primary beam (PB) as a function of time, frequency, and
polarization. The algorithm employs the orthonormal, circular Zernike polynomial
basis to model the full Jones AIP response, which are obtained by a
Fourier transform of corresponding antenna holography measurements.  These full
Jones models are then used to reconstruct the full Mueller AIP response of an
antenna, in principle accounting for all the off-axis frequency dependent
leakage effects of the primary beam.  The A-to-Z solver is general enough to
accommodate any interferometer for which holographic measurements exist, and we
have successfully modelled the AIP of VLA, MeerKAT and ALMA as a demonstration
of its versatility.  We show that our models capture the PB morphology to high
accuracy within the first 2 sidelobes, and show the viability of full Mueller
gridding and deconvolution for any telescope given high quality holographic
measurements.

\end{abstract}

\section{Introduction}

Modern radio interferometers are capable of high sensitivity, high dynamic range
imaging. Imaging in particular is limited by the presence of direction dependent
effects (DDEs).  In general DDEs are a function of direction, frequency, time
and polarization and are typically corrected for during the imaging process
unlike direction independent effects (DIEs). Following
\cite{hamaker1996understanding}, the measurement equation in radio
interferometry is of the form

\begin{align}
  \vect{V}_{ij}^{Obs} = \matr{G}_{ij}\int \matr{M}_{ij}(\vect{s}) \vect{I}_{ij}(\vect{s}) e^{i\vect{b}_{ij}.\vect{s}}d\vect{s}
  \label{eq:meas_eq}
\end{align}

\noindent for a single baseline $i$---$j$ for a given frequency at a given time,
$\matr{G}_{ij}$ are the DIEs, $\matr{M}_{ij}$ are the DDEs, $\vect{I}_{ij}$ is
the sky brightness distribution, $\vect{s}$ defines the direction vector on
the sky, and $\vect{b}_{{ij}}$ are the \textit{uv} coordinates of the baseline
$i$---$j$. All the terms within the integral in the above equation have to
be corrected for during the imaging process, as they are all functions of the
direction vector $\vect{s}$.

$\matr{M}_{ij}$ is a $4 \times 4$ matrix  describing the direction dependent
(DD) mixing of the full-polarization image 4-vector ($I_{pp}$, $I_{pq}$,
$I_{qp}$, $I_{qq}$).  Each element of $\matr{M}_{ij}$ is a description of the
DD response of an interferometer. The diagonal elements represent the
power response (\textit{i.e.,} forward gain) of the interferometer. Elements in
the first row (or column) encode the first order DD polarization leakages due to
antenna optics. Other off-diagonal elements encode higher order combinations of
power and polarization leakage terms.  These are typically order of magnitude
smaller (but not always).  Accurate models for the elements of
$\matr{M}_{ij}$, particularly the leakage terms, is a prerequisite for
full-Mueller imaging that corrects for the effects of the antenna DD response in
full polarization.  This work describes a method to develop a model for
$\matr{M}_{ij}$ based on holographic measurements of the antenna response.

The Mueller matrix $\matr{M}_{ij}$ can be written as
\begin{equation}
  \matr{M}_{ij}(\vect{s}, \nu, t) = \matr{J}_{i}(\vect{s}, \nu, t) \otimes \matr{J}_{j}^{*}(\vect{s}, \nu, t)
\end{equation}

\noindent where $\otimes$ is the Kronecker product, and $\matr{J}_{i}$ and
$\matr{J}_{j}$ are the antenna voltage patterns or the direction dependent Jones
matrices for antennas $i$ and $j$ respectively as a function of frequency $\nu$
and time $t$. The antenna Jones matrix encodes the polarization response of an
antenna to incident radiation and is represented by a $2\times 2$ matrix given
by

\begin{equation}
  \matr{E} = \begin{pmatrix}
    E_p & -E_{p\leftarrow q} \\
    E_{q \leftarrow p} & E_q
    \end{pmatrix}
\end{equation}

\noindent
for two orthogonal feed polarizations $p$ and $q$.

In order to accurately reconstruct the sky brightness distribution it is necessary to remove
the imprint of $\matr{M}_{ij}$ for each baseline for all direction, frequency
and time.

Formally, the PB of an antenna is given by the Fourier transform of the aperture
illumination pattern (AIP) and can be represented as

\begin{equation}
  \matr{A}_i(u,v,\nu,t) = \mathcal{F}\matr{J}_i(\vect{s},\nu,t)
  \label{eq:AIP}
\end{equation}

\noindent where $\matr{A}_i$ is the complex AIP of antenna $i$, at \textit{uv}
coordinates $u$, $v$. $J_i$ is the measured complex image plane Jones matrix,
(\textit{i.e.,} antenna voltage pattern) and $\mathcal{F}$ is the Fourier
transform operator.  $\matr{A}_{i}$ is mathematically a finite, bounded function
and correspondingly $\matr{J}_{i}$ is unbounded. The AIP in general is a
function of time, frequency, and polarization. The time dependence for an
altitude-azimuth mounted antenna manifests as a rotation of the source within
the field of view. We can consider measurements made by a radio interferometer
as the sampling of a continuous visibility coherence function by the AIPs at the
locations of the baselines in the \textit{uv}-plane. The measurement equation
(Eq.~\ref{eq:meas_eq}) can be recast using Eq.~\ref{eq:AIP} (following
\citealt{bhatnagar2013wide}) as

\begin{equation}
\vect{V}^{obs}_{ij}(\textbf{s},\nu,\theta_{PA}) = \matr{A}_{ij}(\textbf{s},\nu,\theta_{PA}) \circledast \vect{V}_{ij}^{True}
 \label{eq:HBS}
\end{equation}

\noindent where $\matr{A}_{ij} = \mathcal{F}\matr{M}_{ij}$ is the AIP (given by
the Fourier transform of Eq.~\ref{eq:meas_eq}), $\circledast$ is the outer
convolution operator (as described in \citealt{bhatnagar2013wide}), $\theta_{PA}$ is
the parallactic angle and $\vect{V}_{ij}^{True} = \mathcal{F}\vect{I}_{ij}(\vect{s})$ is the
continuous true sky coherence function that is sampled by the baseline AIP
$\matr{A}_{ij}$ resulting in the observed visibility data given by
$\vect{V}^{obs}_{ij}$.

There are various algorithms that are used to mitigate the effect of DDEs such
as peeling and facet-based algorithms \citep[e.g.,][]{cotton2004beyond,
noordam2004lofar, intema2014spam, van2016lofar}  and projection algorithms
\citep[e.g.][]{cornwell2008noncoplanar,bhatnagar2008correcting,bhatnagar2013wide,tasse2013applying,van2018image}
both of which require a model of either the PB or the AIP (refer to
\citet{rau2009advances} for a more details).

In this paper, we restrict ourselves to a discussion on PB correction via projection
algorithms, specifically A-projection. However we note that the A-to-Z
solver modeling approach is itself agnostic to the choice of imaging algorithm.
Although we create the models in the aperture (\textit{i.e.,} data) domain, a
Fourier transform is sufficient to provide equivalent models for image domain
algorithms.

The A-projection algorithms applies an \textit{a priori} model inverse of the
antenna AIP ${\matr{A}_{ij}^{M}}^{\dagger}/|\matr{A}_{ij}|^2$ at the time of
convolutional gridding such that

\begin{equation}
  \mathcal{F}^{\dagger}\left[\frac{{\matr{A}_{ij}^{M}}^{\dagger}}{|\matr{A}|^2}\right]*\vect{V}^{obs}_{ij} = \mathcal{F}^{\dagger}\left[\vect{V}_{ij}^{True}\right] = \vect{I}_{ij}
\end{equation}

\noindent The resulting image is free of the time, frequency and polarization
DDEs of the baseline $i-j$. The DDEs are ``projected out'' by using a gridding
convolution function that is equal to the inverse of the model AIP,  thus
recovering the true sky brightness distribution. The quality of the model and
its inverse then determines how well we can correct for the DDE of the baseline
AIP. There have been many different approaches to fitting, modeling, and
simulating the antenna PB response at different observatories
\citep[e.g.,][]{du2016,jagannathan2017directionii,sokolowski2017calibration,asad2021}.
The primary challenge of pure modeling approaches such as electromagnetic
simulations or ray tracing lie with determining the off-diagonal (leakage) Jones
response of the antenna. Capturing these leakage terms is necessary to be able
to perform accurate widefield polarimetric observations. In all cases, and
particularly for dish antennas, measuring the antenna Jones beams via
holographic measurements yields the ``ground truth'' of the antenna Jones response.

We approach the problem in the data domain where we model the bounded and finite
AIP \citep{bates1971holographic,ScottRyle1977}. We also restrict our discussions
to radio interferometers composed of dishes for the rest of the paper,
specifically focusing on the Karl. G. Jansky Very Large Array (VLA)
\citep{perley2011expanded}, Atacama Large Millimeter Array (ALMA)
\citep{wootten2003atacama} and MeerKAT \citep{jonas2018meerkat}.  We present
here a new method, A-to-Z solver, to derive a Zernike polynomial based model of
the AIP from measurements of antenna holography and demonstrate its accuracy and
efficacy in modeling the full Jones response of the antenna.  This allows us in
turn to generate full Mueller AIP and antenna PB models. This paves the way for
a full Mueller treatment of A-projection, which will result in accurate
widefield, wideband off-axis polarimetry. This is currently under development
and will be described in an upcoming paper (Jagannathan et al., \textit{in
prep}). This paper and the next is a part of our effort to implement full
Mueller polarization corrections in the A-Projection algorithm in CASA
\citep{mcmullin2007casa}.

The rest of the paper is organized as follows :
Section \ref{sec:aperture_to_zernike} provides the details of how we go from
Jones measurements in the image domain to Zernike polynomial models of the
aperture.  Section \ref{sec:results} discuss the results of the modeling, the
accuracy and reproducibility etc., and Section \ref{sec:conclusion} provides a
summary of the methods and their merits and limitations.

\begin{figure*}
  \gridline{\fig{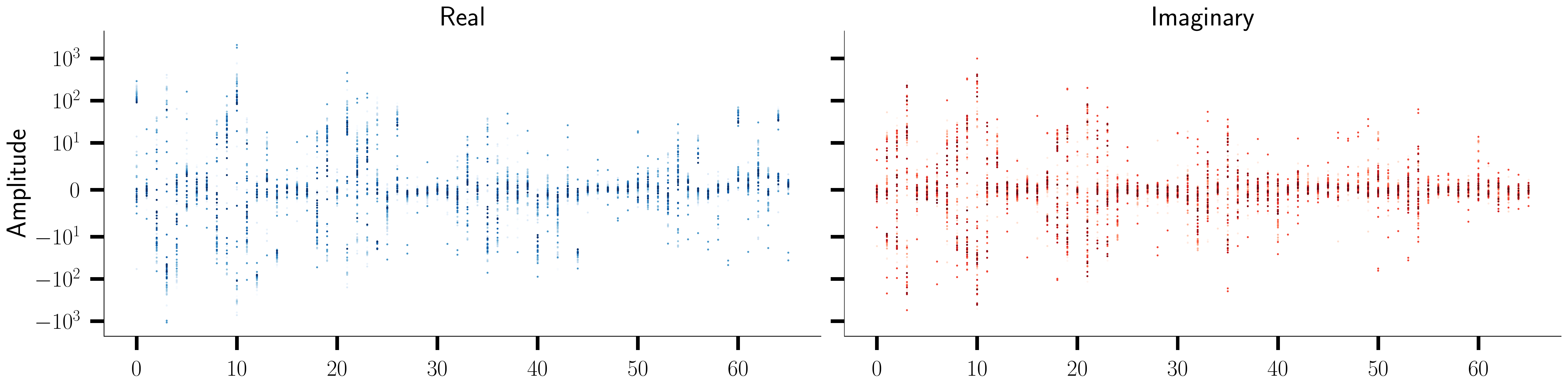}{\textwidth}{(a) EVLA}}
  \gridline{\fig{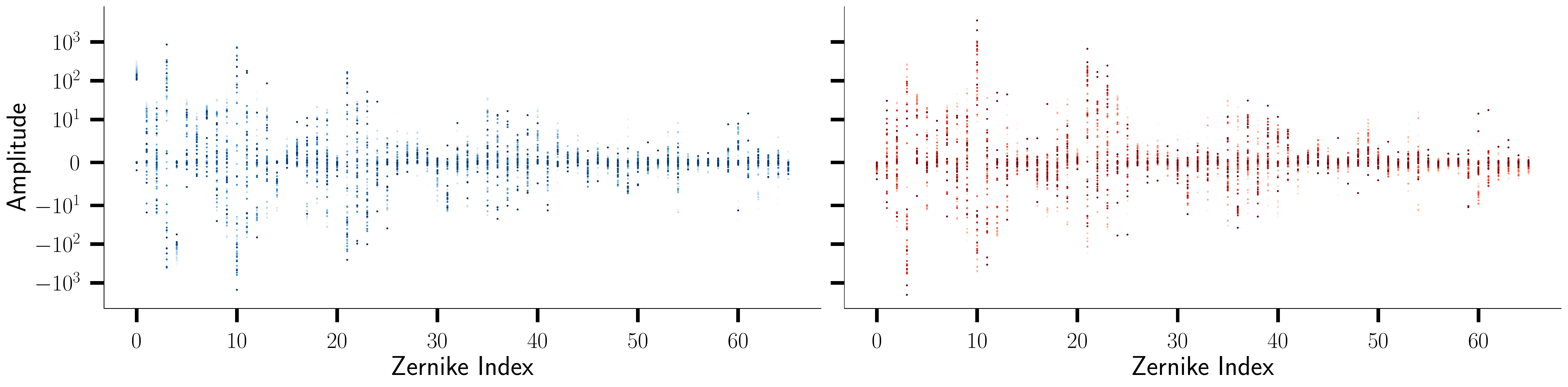}{\textwidth}{(b) MeerKAT}}
  \caption{Plot of the power per Zernike term for the real and imaginary parts
            of the AIP for EVLA (top) and MeerKAT (bottom). The y-axis scale is in arbitrary
            units, since the FFT to convert the Jones beams to apertures scales all the
            pixels by a factor of $\sqrt{N_{pix}}$.  This is has no impact on the accuracy
            of the modeling or reconstruction since the relative power between the different
            Zernike terms is what determines the morphology of the reconstructed aperture,
            and is a conserved quantity under this transformation.}
  \label{fig:zcoeffs_power}
\end{figure*}

\begin{figure*}
  \plottwo{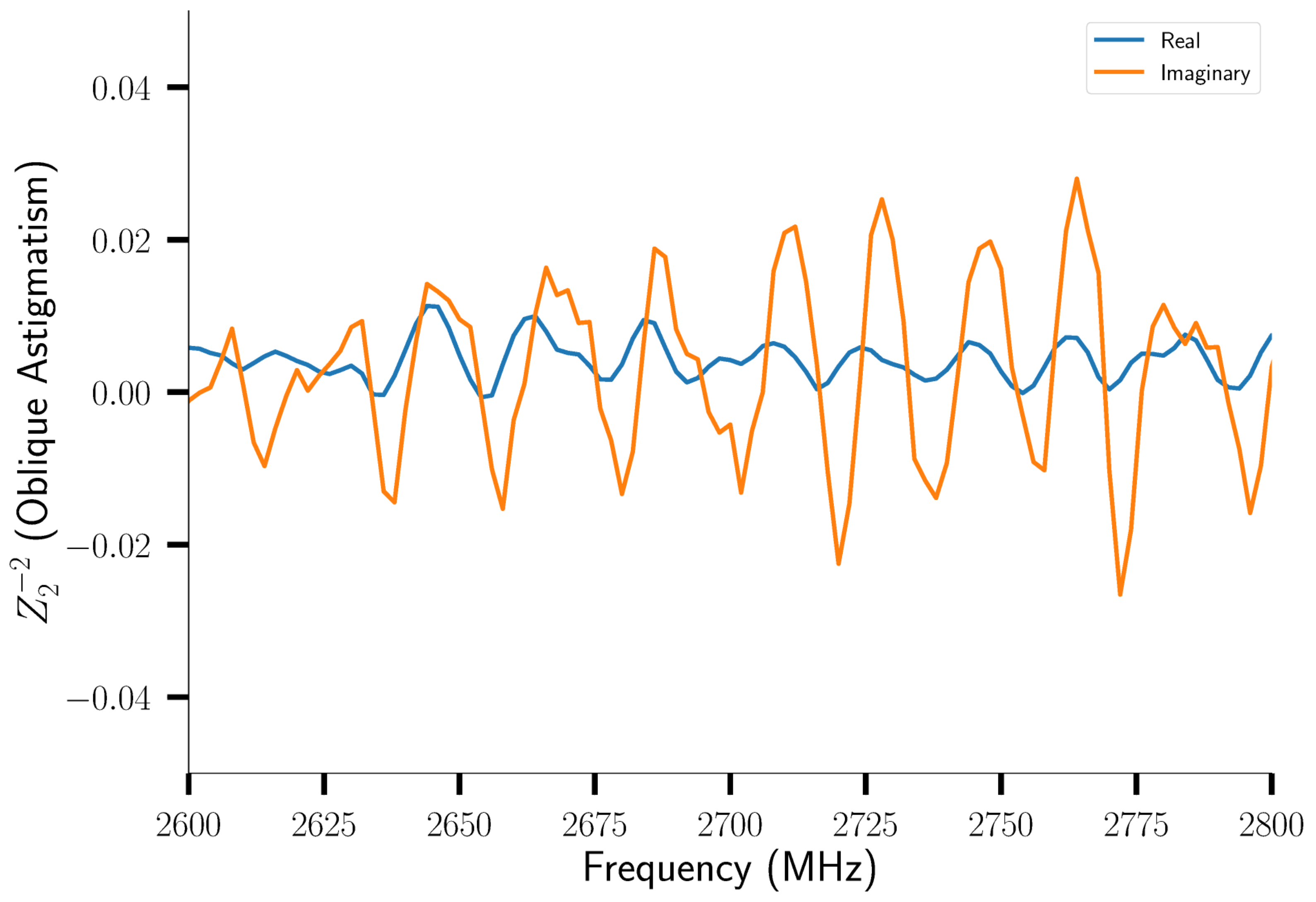}{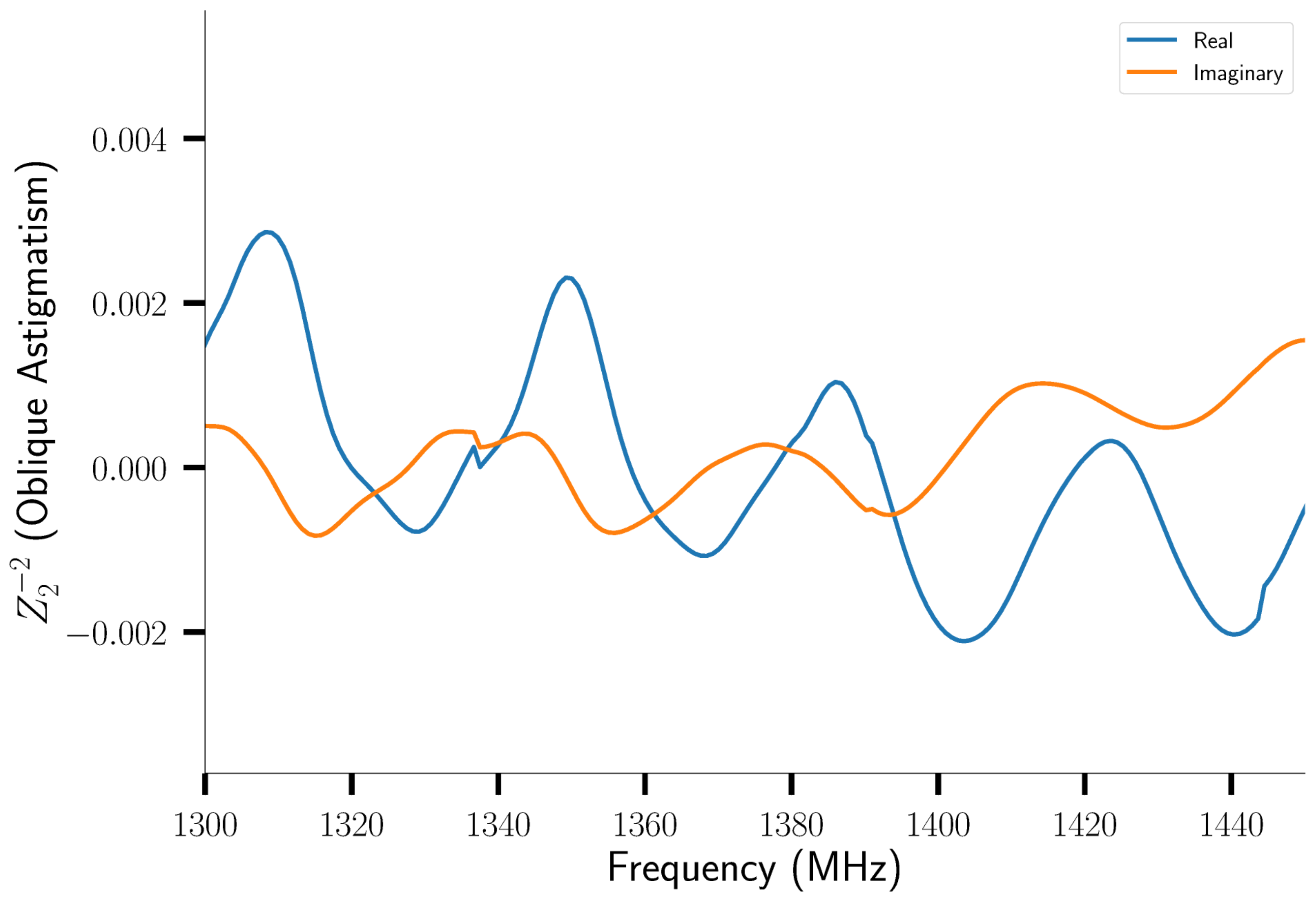}
  \caption{Normalized fitted value of $Z_{2}^{-2}$ (Oblique Astigmatism) from
  VLA and MeerKAT apertures. The sinusoidal patterns present in these
  coefficients correspond to the standing waves in the antenna. \textbf{Left:} The
  coefficient power across a section of the VLA S band, the coefficients capture
  the standing wave of $\sim 17$ MHz. \textbf{Right}: The coefficient power across
  a section of the MeerKAT L band. The standing wave of frequency $\sim 37$ MHz is
  captured. These standing waves correspond to the secondary reflection between
  the antenna surface and the feed.}
  \label{fig:standing_wave}
\end{figure*}

\section{From Apertures to Zernike}
\label{sec:aperture_to_zernike}

\subsection{Interferometric Holography}
\label{sec:holographic_measurements}

Antenna holography is the process of measuring the far-field voltage pattern of
an antenna \citep{bates1971holographic, napier1971holographic}, either by
pointing at a reference (terrestrial) source, or a well characterized celestial
calibrator source. In order to measure the Jones matrices, a two-dimensional
raster scan is done around a known, unresolved, and preferably unpolarized,
calibrator source.  Half the array tracks the calibrator source (the ``tracking
antennas'') while the other half performs the raster scan (the ``scanning
antennas''). The two halves are then swapped around, in order to get a
measurement of all the antennas in the array.

We can recast Equation \ref{eq:meas_eq} in the following form (in vector
notation) under the assumption that all the tracking antennas are pointing at
the same location on the sky as

\begin{equation}
   \vect{V}_{st} = \matr{G}_{st} \mathcal{F}\left[(\matr{J}_{s} \otimes \matr{J}_{t}^{\dag}) \matr{S} \vect{I}^{sky}\right]
\label{eq:holo}
\end{equation}

\noindent where the subscripts $s$ and $t$ refer to the scanning and
tracking antennas respectively. $\matr{G}_{st}$ are the direction independent
gains and $\matr{S}$ is a transformation matrix that converts the sky brightness
distribution from Stokes basis into the feed basis of measurement. The Jones
matrices $\matr{J}_{s}$ and $\matr{J}_{t}$ represent the antenna far field
voltage patterns in feed basis of the scanning and the tracking antennas.  After
calibration of the tracking antennas is performed (using the point source
calibrator), the above equation reduces to

\begin{equation}
  \vect{V}_{st}^{cal} = \mathcal{F}\left[(\matr{J}_{s} \otimes \mathds{1}) \right]
\label{eq:holo_cal}
\end{equation}

\noindent where the $\matr{J}_t$ is reduced to the identity matrix $\mathds{1}$,
and we can therefore measure the complex Jones response (as a function of
direction) of the scanning antennas from the calibrated visibilities. The
product $\matr{S}\vect{I}^{sky}$ is the calibrator source flux as measured in
the feed basis. Since the calibrator flux is nominally well known, this becomes
a constant (known) factor that can be normalized out.

Further, correlation products from all the pairs of scanning-scanning antennas
can be constructed, given by:

\begin{equation}
   \vect{V}_{ss} = \matr{G}_{ss} \mathcal{F}\left[(\matr{J}_{si} \otimes \matr{J}_{sj}^{\dag}) {S} \vect{I}^{sky}\right]
\label{eq:total_power_holo}
\end{equation}

\noindent where $\matr{J}_{si}$ and $\matr{J}_{sj}$ are Jones beams
corresponding to the baseline $i$---$j$. $\vect{V}_{ss}$ then measures the total power
Jones beam. These total power measurements measure the first row of the Mueller
matrix. It is not possible to reconstruct the antenna Jones matrix from these
Mueller measurements. However, the first row of the Mueller matrix can be
reconstructed from the Jones beams (Eq.~\ref{eq:holo_cal}), and provide an
internal consistency check for our measurements. We describe below the details
of the holography for three different instruments (VLA, MeerKAT, and ALMA).

\paragraph{VLA} We use the holography data described in \cite{perley2016evla}.
In this paper, we discuss only the S-Band observations, although the method is
applicable to any observable frequency band at the VLA. At S Band the
observations used 3C147 as the standard unpolarized calibrator. The scanning
antennas covered a regular grid of 57x57 points around the calibrator, with an
angular separation of $2.62^{\prime}$ between each pointing. This was sufficient
to sample out to 5 sidelobes at the highest end of the band. Two spectral
windows (centred at 2308 MHz and 2948 MHz) were badly affected by persistent
RFI, and we could not get reasonable beam measurements from them. We instead
utilize ray traced antenna AIPs which were modelled identically to the rest of
the band to derive the Zernike polynomial coefficients.

\paragraph{MeerKAT} We used MeerKAT L band holography data that covered the
entire MeerKAT L band (from $\sim$ 880 to 1680 MHz) at a resolution of
$\sim 0.85$ MHz. The data sampled out to 5 sidelobes at the highest frequency.
The holographic pointing was performed in a spiral pattern around the calibrator source
3C273, utilizing the on-the-fly (OTF) capabilities of MeerKAT. The beams were
then resampled on to a regular $128 \times 128$ with
a separation of 4.68 arcmin per grid pointing. For more details please refer to
\citet{asad2021}.

\paragraph{ALMA} We use ALMA holography data described in
\cite{bhatnagar2020alma}, obtained in 2018. We focus here only on the Band 3
data, which measured out to $\sim 5$ sidelobes of the PB. The holography sampled
a 49x49 grid with a spacing of 0.2$\times$HPBW (half power beam width) around
the calibrator source J1924-2914. The full Jones beams for both the 12m
antennas, DA and DV, were recorded. The DA and DV antennas are two of the
  three types of 12m antennas that constitute ALMA. The primary difference between
  the antennas is the position of the antenna feed legs, which are rotated by $45^{\circ}$.

\subsection{Pointing Offset Correction}

Prior to modeling the measured aperture, any residual pointing errors in the
holography need to be removed first, as small pointing errors cause a phase
gradient across the aperture \citep{bhatnagar2017pointing}. If these phases
remain during the modeling step, they will be captured by the Zernike model
leading to offsets in the generated model PB. While all the telescopes discussed
here have some form of \textit{a priori} models for pointing offsets (and in
some cases perform a dedicated pointing calibration scan), residual pointing
errors tend to accumulate through the course of an observation.

In order to measure and correct for the pointing errors, a 2D Gaussian is fitted
to the main lobe of the holographic beams. We use a non-linear
Levenberg-Marquardt least squares fitter \citep{more1978levenberg} as
implemented in the \texttt{astropy} Python package
\citep{robitaille2013astropy}.

The beam for each feed polarization was fitted independently, and the vector sum
of the pointing vectors between the two feeds is taken to be the pointing
offset. This procedure is repeated as a function of antenna and frequency. We
use the vector sum to preserve the beam squint between the two orthogonal
polarizations. The measured pointing offsets were used to regrid the beam such
that the peak of the Stokes I beam lies at the centre of the image.  All four
Jones beams are regridded identically. For the VLA and ALMA, where we had access
to the holography visibility data, we were able to remove the pointing offsets
per baseline prior to averaging all the baselines of an antenna. This results in
higher SNR upon averaging, eliminating smearing and decorrelation due to
baseline-based pointing errors.

Finally, the holography data were averaged across all antennas and all channels
within a spectral window prior to applying the Fourier transform. This averaging
results in an improvement of  $\sqrt{N_{ant}\times N_{chan}}$ in the SNR which
is especially useful in the cross hands which usually have $\sim 10-100\times$
smaller SNR than the parallel hands. Empirically we have determined that the
average antenna model is sufficient for our modeling purposes. It is however
worth noting that the choice to average across antennas and frequency will
impose a dynamic range limitations on the final image, contingent on the level
of antenna-to-antenna variations and variation across a single spectral window
for a particular instrument.

\subsection{Obtaining the AIP}
\label{sec:obtaining_aip}

The spatial frequency resolution in the aperture domain is inversely
proportional to the total angular extent sampled by holography.  It is given by
:

\begin{equation}
\label{eq:delta_u_delta_l}
  \Delta u \propto \frac{1}{l_{sky}}
\end{equation}

\noindent
where $l_{sky}$ is the total angular extent of the hologaphy (in radians)
and $\Delta u$ is the size of the resolution
element in the aperture (in lambda).

It is necessary that a large number of sidelobes be covered by holography in
order to obtain accurate aperture plane measurements. However in practice only a
finite number of sidelobes can be measured, due to various practical
considerations. In such a case, the corresponding aperture domain measurements
are affected by ringing and these ringing artefacts extend far beyond the
physical aperture.  We therefore need to determine the ``true'' edge of the
aperture in order to model the AIP rather than the ringing artefacts.

The real part of the AIP for an unblocked aperture is guaranteed to be positive
within the physical aperture itself. In the case of a blocked aperture
$\Re(\matr{A})$ can have null or negative values within the regions that are
shadowed. However in both cases, the last pixel inside the aperture (in an
unblocked region) will be positive and the first pixel outside the aperture will
be negative. This is commonly called the ``aperture roll-off''. In absolute
value, the drop between these two pixels can be a factor of between 10 to 100.
At low frequencies, the magnitude of the drop is smaller since there is more
diffraction and spill-over at the edge of the aperture.  The above condition has
proven to be very robust in determining the cutoff radius of the aperture for a
variety of different antenna types and frequencies.  We identify the size of the
aperture by locating the radius of the first negative component in the real part
of the aperture, in a direction moving out from the centre of the aperture. This
condition is general, and allows us to naturally determine the cutoff radius as
a function of frequency.

The apertures derived from antenna holography have a gradual rolloff at the edge
of the aperture, due to the limited number of sidelobes that can be sampled by
holography. The diameter of the main lobe of the PB depends very strongly on the
size of the aperture, and therefore in order to get the correct beam sizes it is
necessary to determine the aperture size as accurately as possible.
Directly applying a Fourier transform to the measured holography typically
results in 5-6 pixels across the aperture, and the roll-off is contained in a
single pixel. This can bake in an error of up to 15\% in the derived PB size.

In order to obtain accurate ($< 1$\% error) PB sizes we used an oversampling
factor of 100 prior to the Fourier transform of the holography followed
by image plane minimization (see Sec.~\ref{sec:aperture_eta}) to account for the changing aperture efficiency
across the band. We performed this interpolation in CASA using the
\texttt{imregrid}\footnote{https://casa.nrao.edu/casadocs/casa-6.1.0/global-task-list/task\_imregrid/about}
task which performs cubic spline interpolation, while preserving pixel flux
scaling.  We note here that interpolation of any kind does not increase the
amount of information in the underlying image. Therefore interpolating by a
large factor, while allowing us to determine the edge of the roll-off more
accurately, does not increase the steepness of the roll-off or give us any more
resolution across other features in the aperture.

\subsection{Aperture Fitting}
\label{sec:aperture_fitting}

\begin{figure*}
  \gridline{\fig{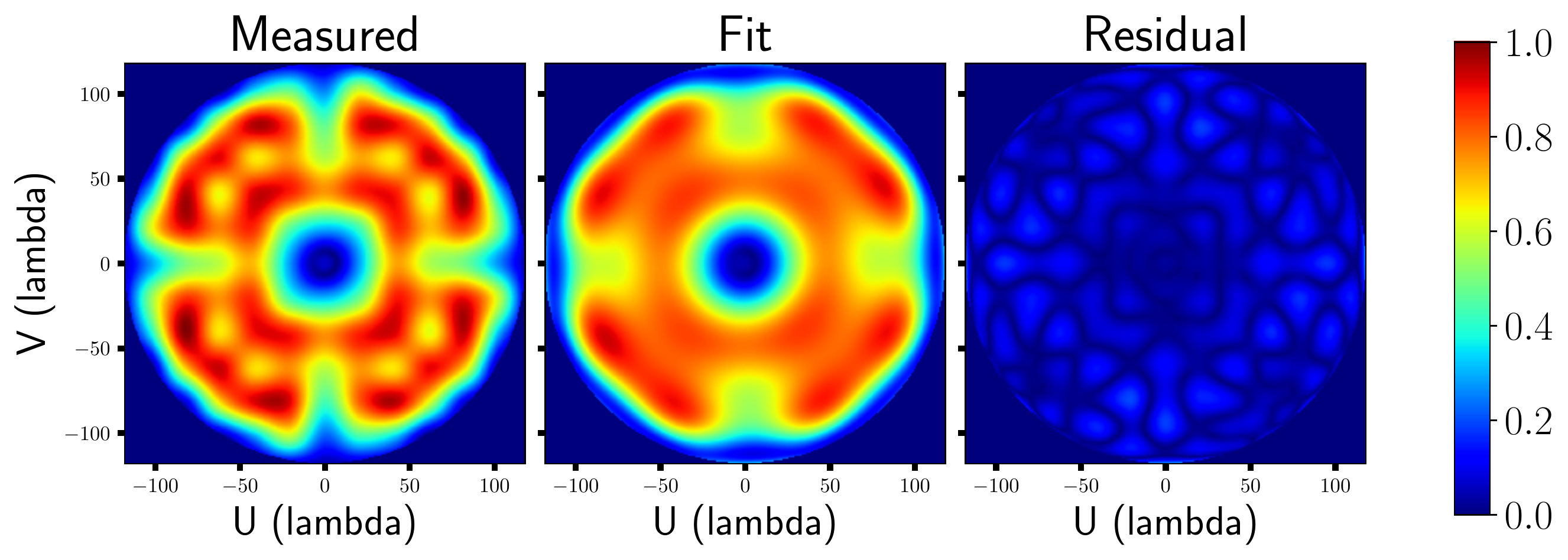}{0.5\textwidth}{(a) EVLA}
            \fig{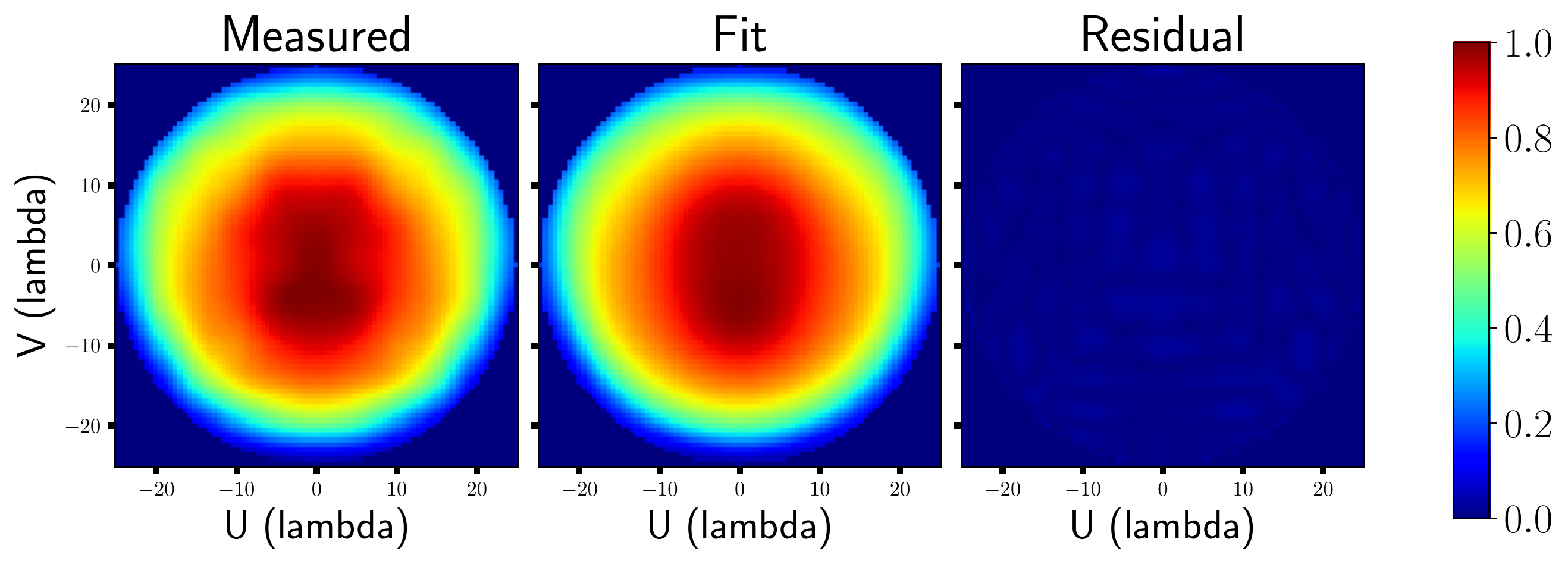}{0.5\textwidth}{(b) MeerKAT}}
  \gridline{\fig{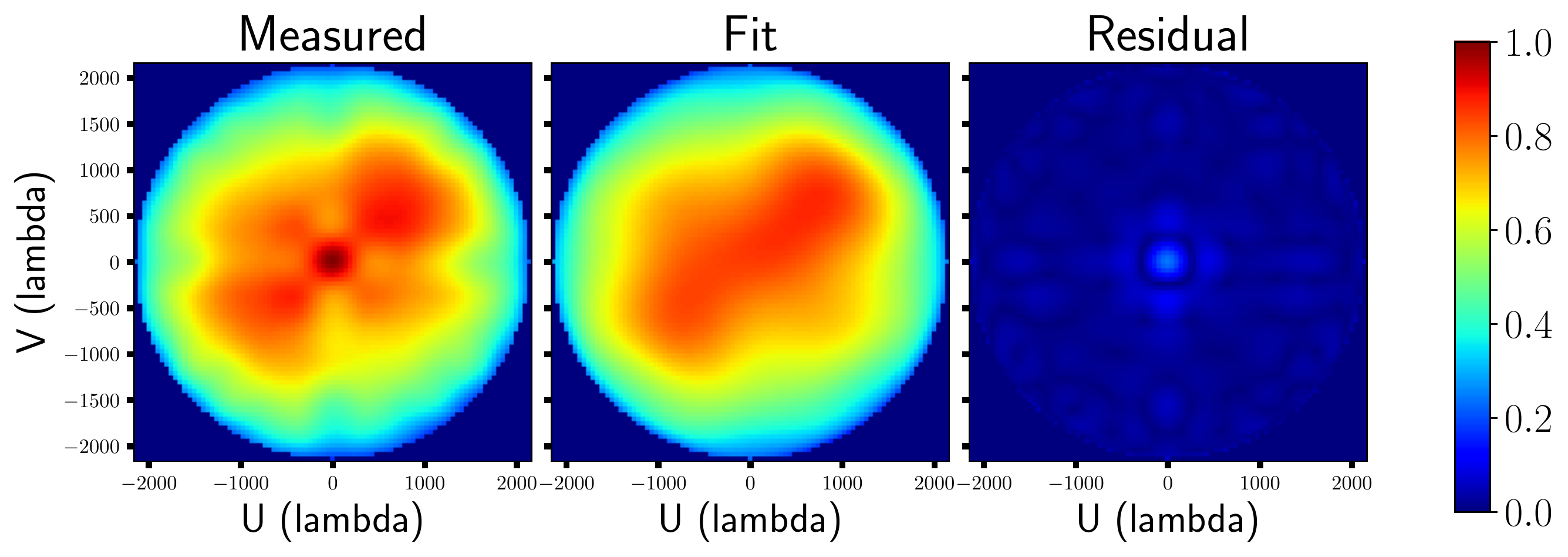}{0.5\textwidth}{(c) ALMA DV}
            \fig{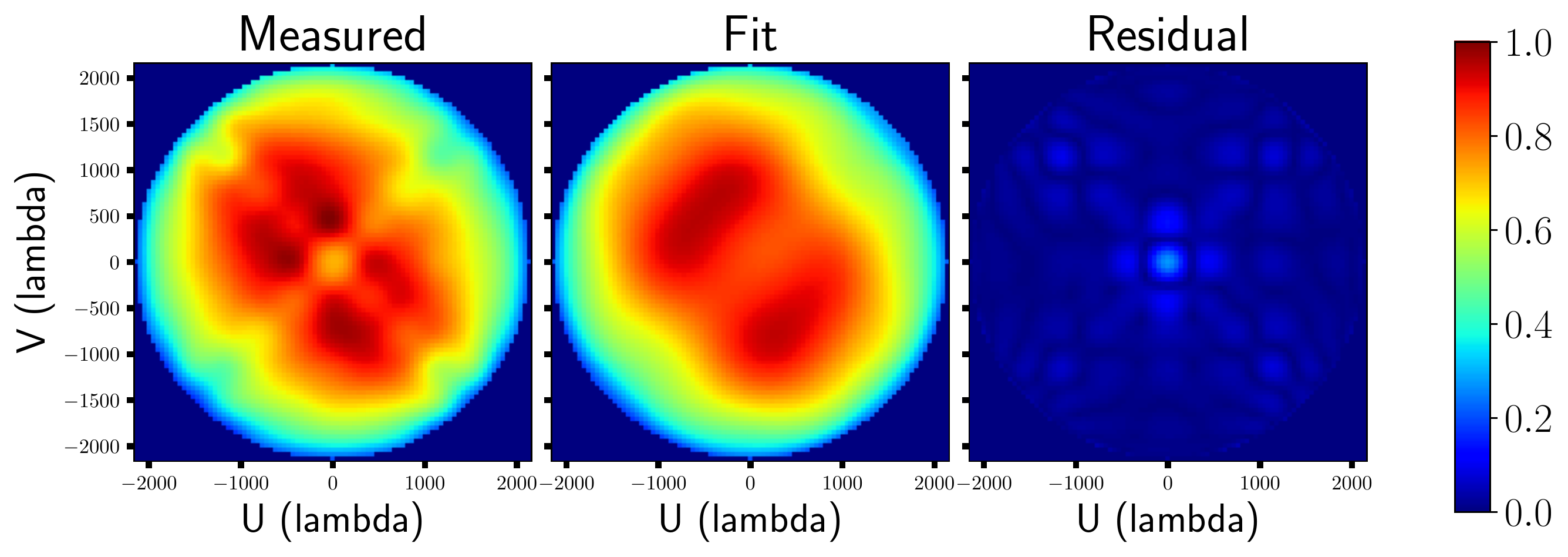}{0.5\textwidth}{(d) ALMA DA}}

          \caption{The measured, model, and residual apertures for each of the 4
          antenna types discussed in this paper. In all cases, only the amplitude term for
          a single polarization is shown, for brevity. The residuals look similar across
          polarization in all cases. (a) VLA S Band (3 GHz) R polarization (b)
          MeerKAT L Band (1.2 GHz) X polarization (c) ALMA Band 3 (108 GHz) DV type X
          polarization, (d) ALMA Band 3 (108 GHz) DA type X polarization. The apertures
          have been normalized to the peak illumination value. In every case the residuals
          show systematic higher order structure, that correspond to power in the higher
          sidelobes. With the current level of modeling we are able to capture all the
          power in the main lobe and the first sidelobe.}

  \label{fig:aperture_fits}
  \end{figure*}

Zernike polynomials are complex, ortho-normal polynomials defined on a unit circle
\citep{zernike1934,born2013principles}. Zernike polynomials also form the
natural basis to model optical apertures, making them an ideal fit for modeling
the antenna AIP.

We refer to \citet{lakshminarayanan2011zernike} for the definition of the
Zernike polynomials, however we use Noll sequential indices
\citep{noll1976zernike} to map the two Zernike indices $(n, m)$ to a single
index $k$. Under this mapping, the first 10 orders ($n : 0 \rightarrow 10$) of
Zernike polynomials map to the first 66 terms ($k : 0 \rightarrow 66$) of the flattened index.

We fit the aperture measurements with the first 10 orders (\textit{i.e.,} 66
terms) of Zernike polynomials after first taking out the known systematic errors
from the holography image (such as antenna pointing offsets).

The AIP model is given by

\begin{equation}
\matr{A}^M(\vect{s}) = \sum_{k=0}^{10} \left[\matr{A}^{\Re}_k + \iota
  \matr{A}^{\Im}_k\right] \cdot Z(k,\vect{s})
\end{equation}

\noindent where $\vect{s}$ spans the aperture of a fixed diameter and $\matr{A}_k$ are
the coefficients for the $k^{th}$ Zernike term $Z(k,\vect{s})$.  The
superscripts $\Re$ and $\Im$ indicate separate coefficients
for the real and imaginary parts of the AIP.   The
objective functions used for deriving the coefficients $\matr{A}^{\Re}$ and
$\matr{A}^{\Im}$ are

\begin{subequations}
\begin{align}
\chi_{\Re}^2 &= \sum_{\vect{s}} \left| \Re\left[\matr{A}^{H}(\vect{s}) - \matr{A}^M(\vect{s})\right] \right|^2\\
\chi_{\Im}^{2} &= \sum_{\vect{s}} \left|\Im\left[\matr{A}^{H}(\vect{s}) - \matr{A}^M(\vect{s})\right] \right|^2
\end{align}
\label{eq:objective_function}
\end{subequations}

where $\matr{A}^{H}$ is the measured AIP from holography observations. The models are
fitted in the aperture domain independently for the real and imaginary parts and
for each of the four Jones terms.  Modeling all the terms of the complex antenna
Jones matrix allows us to reconstruct the entire $4\times 4$ Mueller matrix. The
reverse operation of going from Mueller matrix to Jones matrices is not possible
due to missing phase information in the measured Mueller matrix.  This is the
primary reason for the measurement of antenna Jones matrices (voltage beams).

We use a non-linear least-squares fitter as implemented in the SciPy package
\citep{virtanen2020scipy} to minimize the objective functions defined in
Eq.~\ref{eq:objective_function}. The fitter uses a non-linear trust region
solver algorithm \citep{branch1999subspace} with a stopping threshold criterion.
We define the stopping threshold to be equal to the thermal noise per pointing
in order to prevent over-fitting.
We will demonstrate in the following sections that the correlated nature of our
residuals, albeit small, is a sign of un-modelled zernike terms. We eschew using higher
orders in favor of accurately modeling the main lobe of the antenna PB and the first
sidelobe.

\begin{figure*}
  \plotone{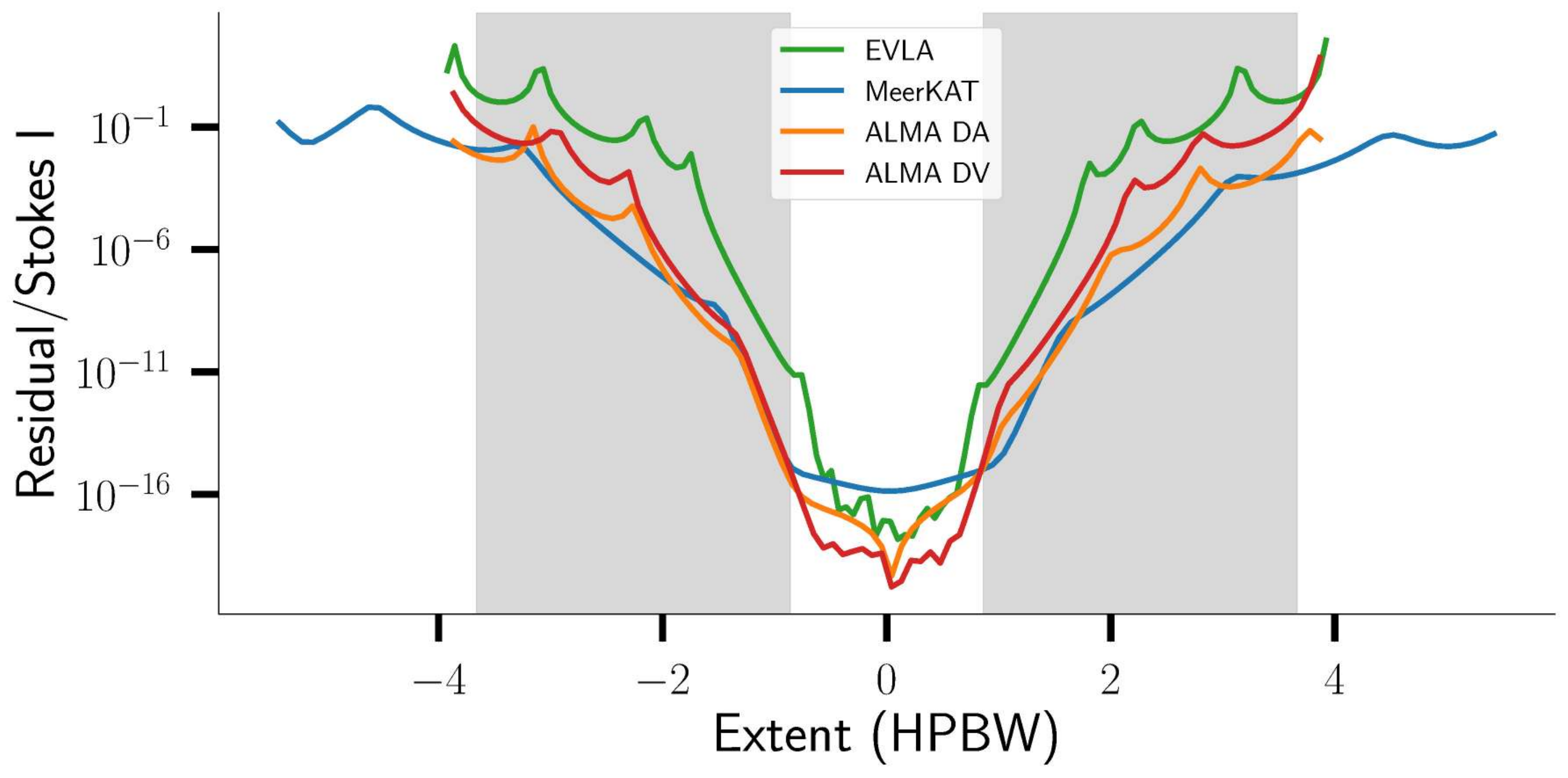}
  \caption{Cut across the normalized residual PB for each of the four antenna
    types discussed here. The plot shows where the un-modelled structure in the
    aperture appears in the image plane.  The MeerKAT beam is at 1.6 GHz, VLA S Band
    at 4 GHz, and the ALMA beams at 108 GHz. In the case of VLA and MeerKAT
    we selected the highest SPW in order to show the largest number of sidelobes.
    The shaded region spans the region from the first null through the fourth
    null, \textit{i.e.,} the first three sidelobes for the EVLA. }
  \label{fig:pb_resid_cut}
\end{figure*}

The plots in Fig.~\ref{fig:zcoeffs_power} show the fitted power per Zernike term
(real and imaginary) for VLA and MeerKAT. The y-scale on the plots is in
arbitrary units.  The distribution of power across the Zernike terms reflects
the different characteristics of the two apertures. The VLA has a blocked
aperture and feed legs which cast a shadow on the dish. Modeling the relatively
sharp edges of the shadow on the aperture requires higher order Zernike terms.
Accordingly, the high spatial frequency Zernike terms (above index 50) see an
increase in power for the case of VLA as seen in Fig~\ref{fig:zcoeffs_power}. In
contrast, MeerKAT has an unblocked aperture and correspondingly has lower
amplitudes for the higher order terms.

As mentioned previously, the Zernike polynomials capture various physically
meaningful optical properties of the aperture 
capture the piston, tip and tilt respectively).  Fig.~\ref{fig:standing_wave}
plots the $Z_{2}^{-2}$ term, or the ``oblique astigmatism'' across a section of
the band for VLA and MeerKAT. This term clearly captures the resonant frequency
dependent variation between the feed and the antenna surface for both the
telescopes. We measure the VLA standing wave of frequency $\sim 17$ MHz,
corresponding to a sub-reflector at a height of $\sim 8.5$m consistent with
known values for VLA \cite[e.g.,][]{jagannathan2017directionii}. For MeerKAT we
measure a standing wave of frequency $\sim 37$ MHz corresponding to a
sub-reflector height of $\sim 4$m which is also consistent with the MeerKAT
specification \citep{esterhuyse2011ska}.

Fig.~\ref{fig:aperture_fits} shows a plot of the measured aperture, the Zernike
model, and the residual for (a) VLA, (b) MeerKAT, (c) ALMA DV and (d) ALMA DA
respectively. The plots show the magnitude (\textit{i.e.,}
$\sqrt{\Re(\matr{A}^{M})^{2} + \Im(\matr{A}^{M})^{2}}$ following
Eq.~\ref{eq:objective_function}) of the aperture for a single polarization and
frequency for each telescope type. As noted earlier the fitting is
performed independently for the real and imaginary components.
The broad morphology of the aperture is captured by the model, leaving behind
residuals at the percent level or less in all cases. The residuals correspond to
the un-modelled higher spatial frequency signal and look similar across
frequency and polarization for all the telescopes.

Fig.~\ref{fig:pb_resid_cut} shows a slice across the residuals of the different
antenna types discussed in this paper. The shaded region spans the region from
the first null to the fourth null, \textit{i.e.,} the first three sidelobes for
the EVLA. These plots are the Fourier transform of the aperture residuals
\textit{i.e.,} $\mathcal{F}\left[\matr{A}^{Holo} - \matr{A}^{M}\right]$ , since
the modeling is performed in the aperture domain. The beam models capture the
response to floating point precision ($10^{-6}$) out to the second sidelobe for
all the modelled telescopes. The residual error rapidly increases as we move
further out, which is expected. From Fig.~\ref{fig:pb_resid_cut} the residuals
correspond to un-modelled power in the outer sidelobes. The higher order
sidelobes correspond to the high spatial frequency features in the aperture
domain, which correspondingly require higher order Zernike polynomials to model.
Empirically, we see that increasing the number of terms used to model the
aperture results in a reduction in the systematics of the aperture plane
residuals.

\section{Results \& Beam Properties}
\label{sec:results}

\subsection{Beam Spectral Index}
\begin{figure*}
  \gridline{\fig{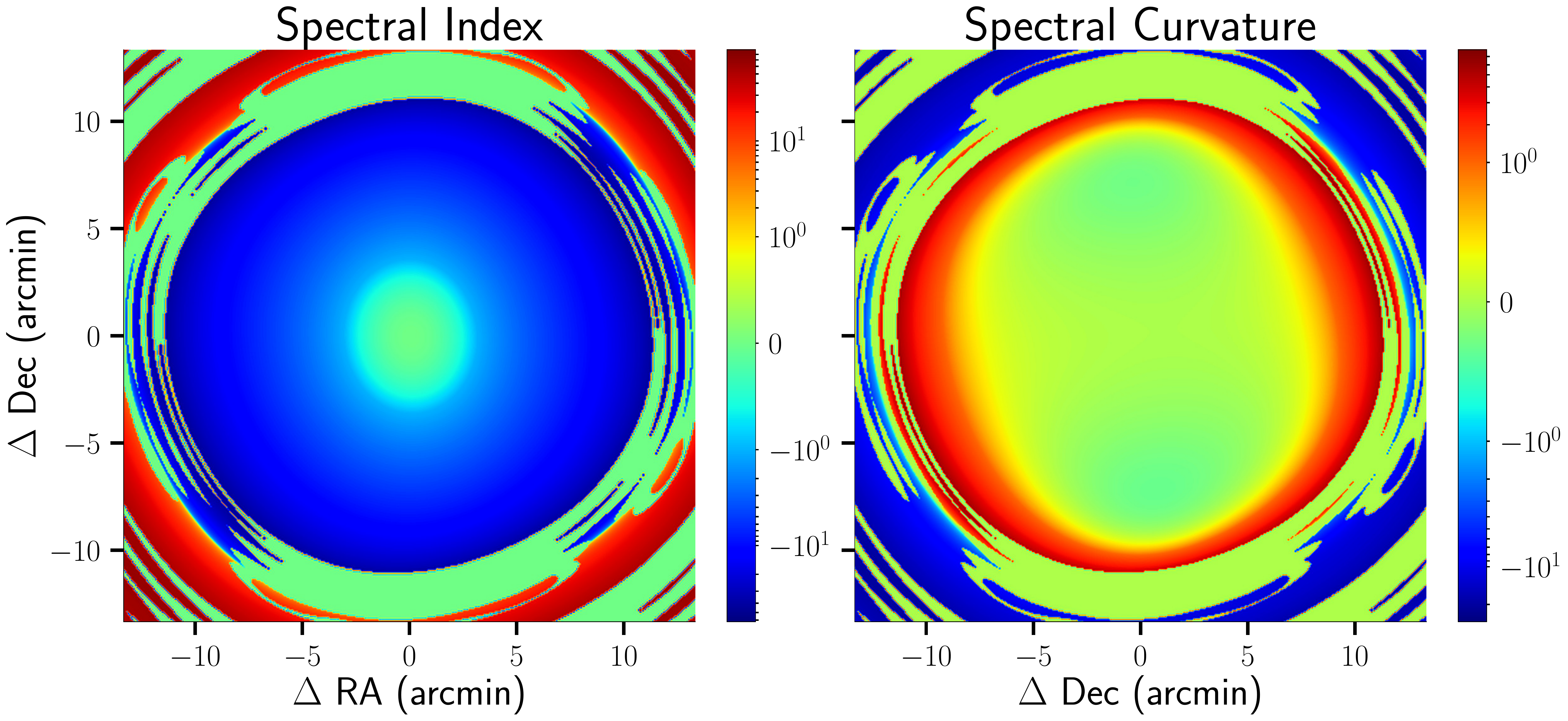}{0.5\textwidth}{(a) EVLA}
            \fig{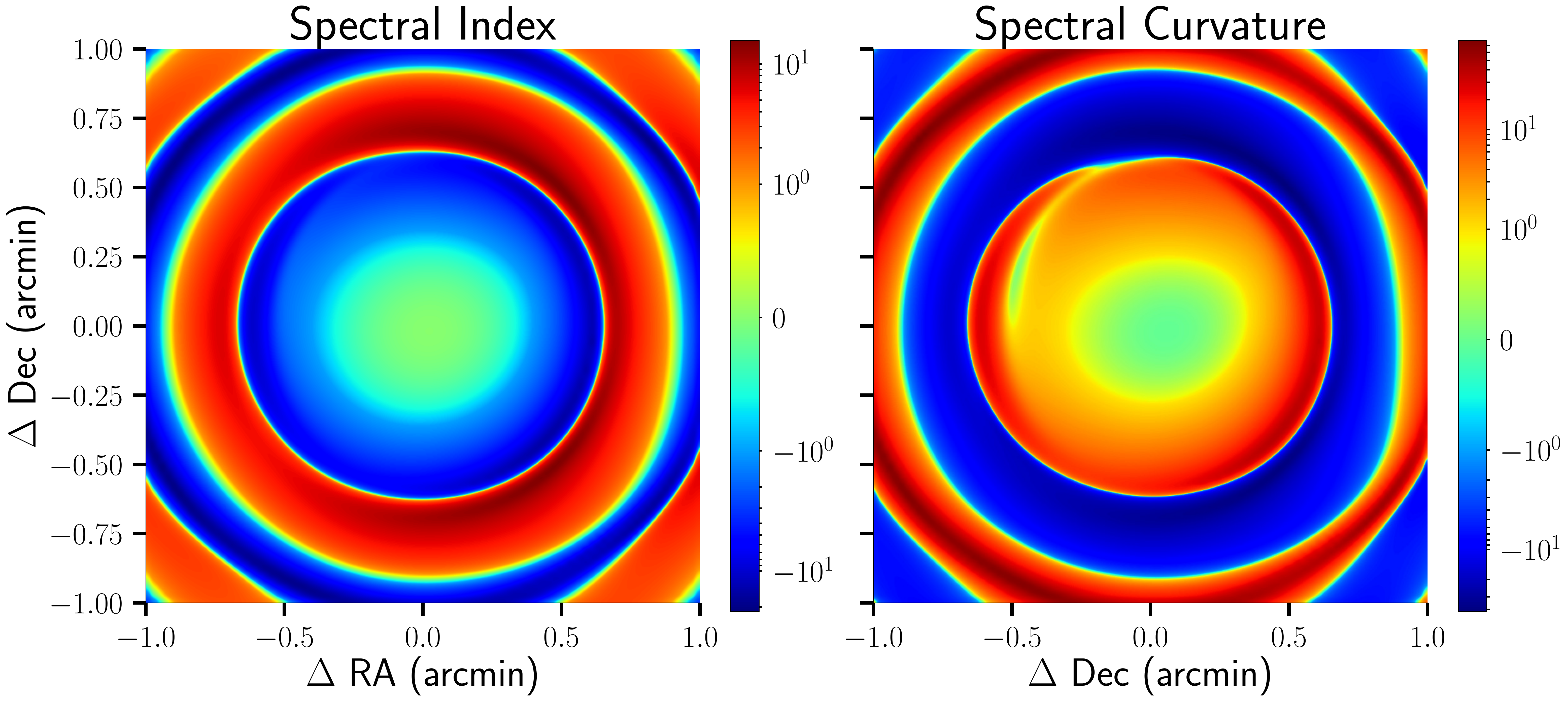}{0.5\textwidth}{(a) MeerKAT}}
  \caption{Plot of the spectral index and spectral curvature of the EVLA S Band (left)
  and MeerKAT L Band (right) beams. The plots show only the main lobes of the
  primary beam with the spectral index steepening as we move further away from the pointing center.}
  \label{fig:pb_spix}
\end{figure*}

With modern interferometers, wideband continuum imaging has become the norm
enabling theoretically  higher sensitivies. In order to achieve these
sensitivity limits it is vital to account for the effects of the wideband antenna
primary beam. Fig~\ref{fig:pb_spix} shows the spectral index $\alpha_{pb}$ and
the spectral curvature $\beta_{pb}$ introduced by the antenna PB onto a wideband
continuum image at~$3$GHz at the VLA on the left panel and at ~$1.2$GHz of
MeerKAT on the right. The spectral index and curvature are defined as

\begin{equation}
  \vect{I}_{\nu} = \vect{I}_{\nu_0} \left(\frac{\nu}{\nu_0}\right)^{\alpha + \beta \log\left(\frac{\nu}{\nu_0}\right)}
  \label{eq:spix_eqn}
\end{equation}

\noindent where $\alpha$ and $\beta$ are the spectral index and spectral
curvature respectively, $\vect{I}$ is the source flux density at frequency $\nu$ and
$\nu_{0}$ is the reference frequency.

The uncorrected PB spectral index at half power of the VLA
PB is ~$\alpha_{pb} = -5$. The various means of mitigating the PB spectral index
have been analyzed in detail in \cite{rau2016}. It is worth noting that for
large continuum frequency bands of observations the higher order spectral terms
of the PB such as curvature also leave a significant spectral signature on the
PB across wide fields of view.

\subsection{Beam Squint \& Squash}
\begin{figure*}
  \gridline{
    \fig{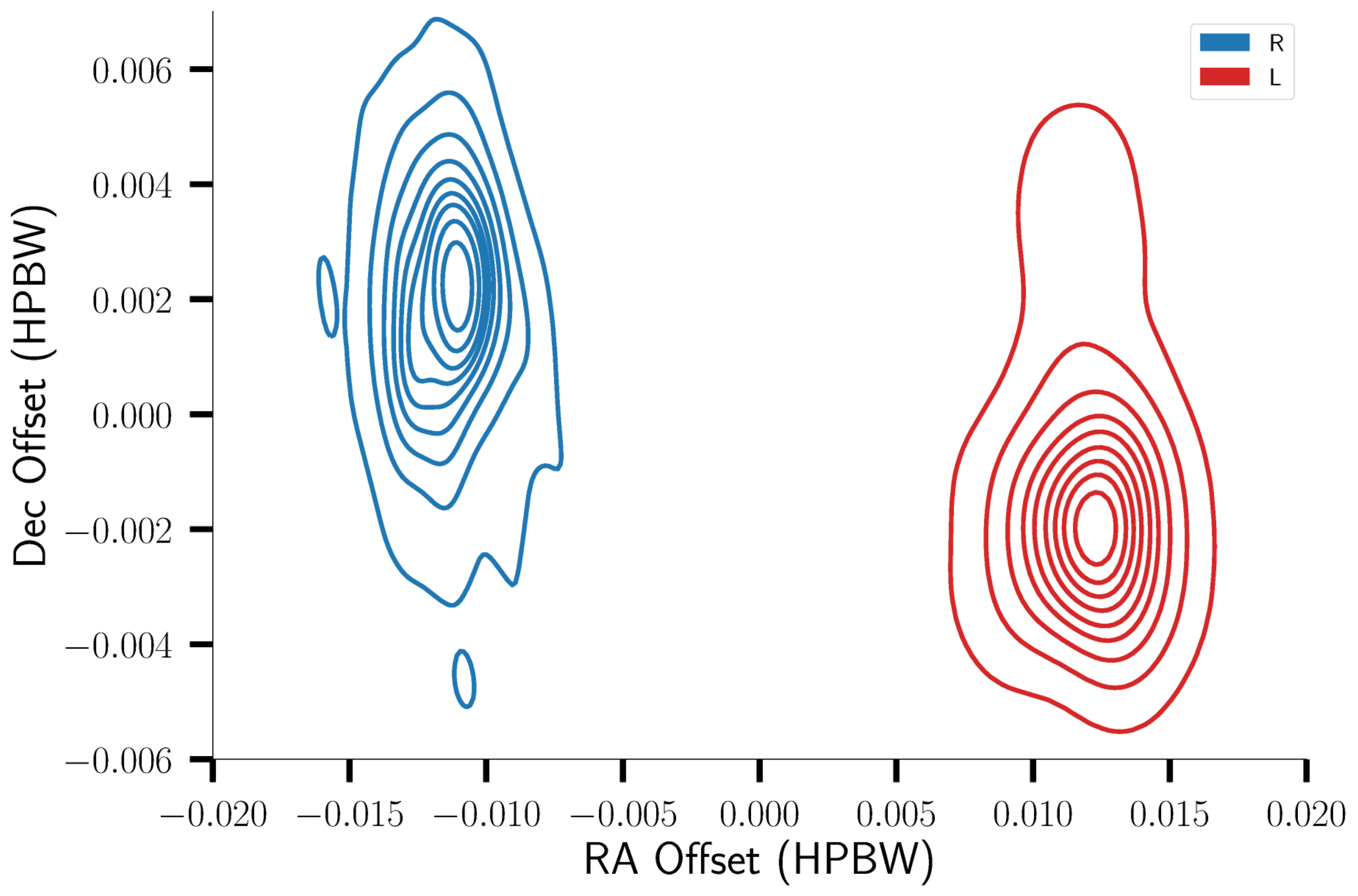}{0.5\textwidth}{(a) EVLA}
    \fig{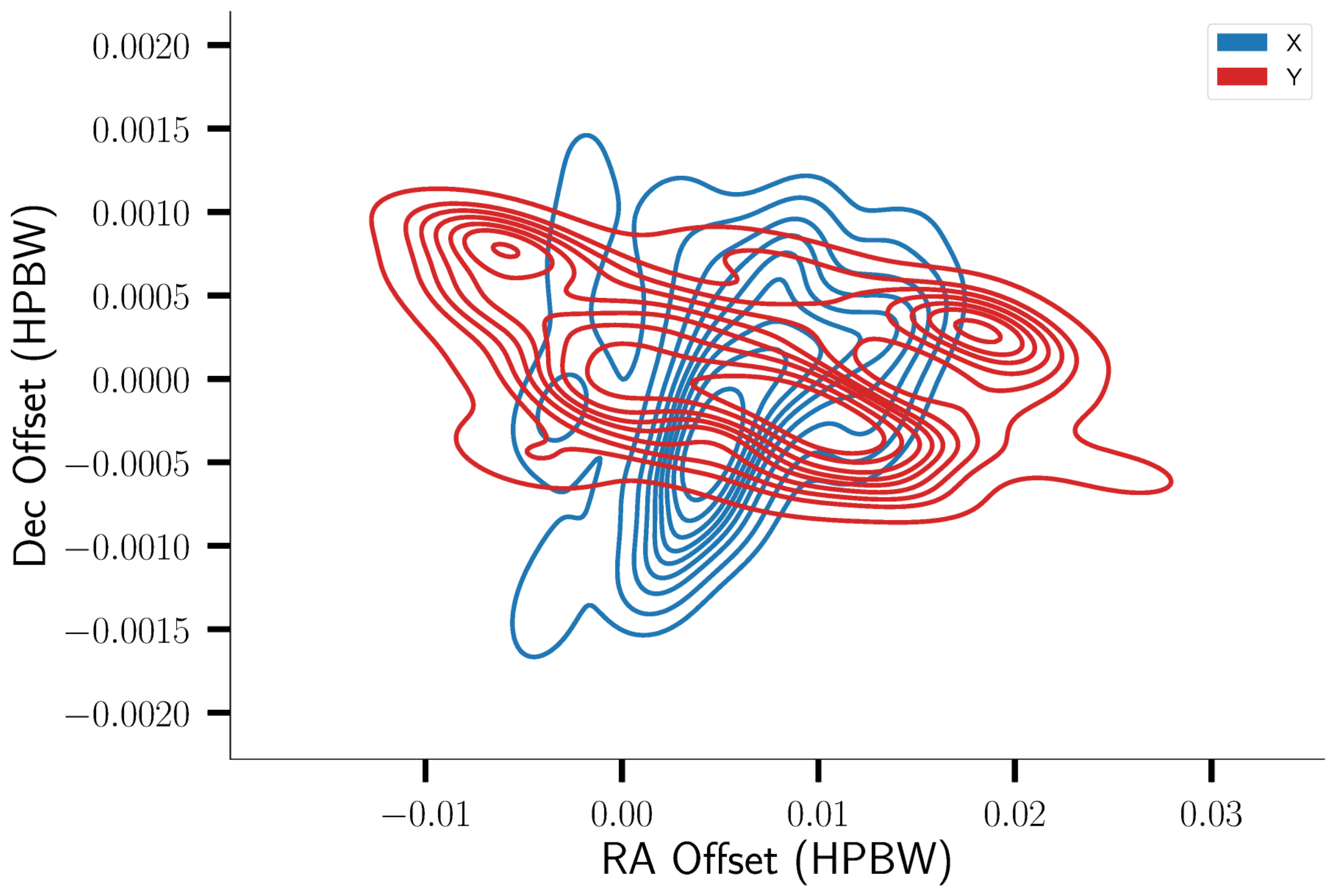}{0.5\textwidth}{(b) MeerKAT}
  }
  \gridline{
    \fig{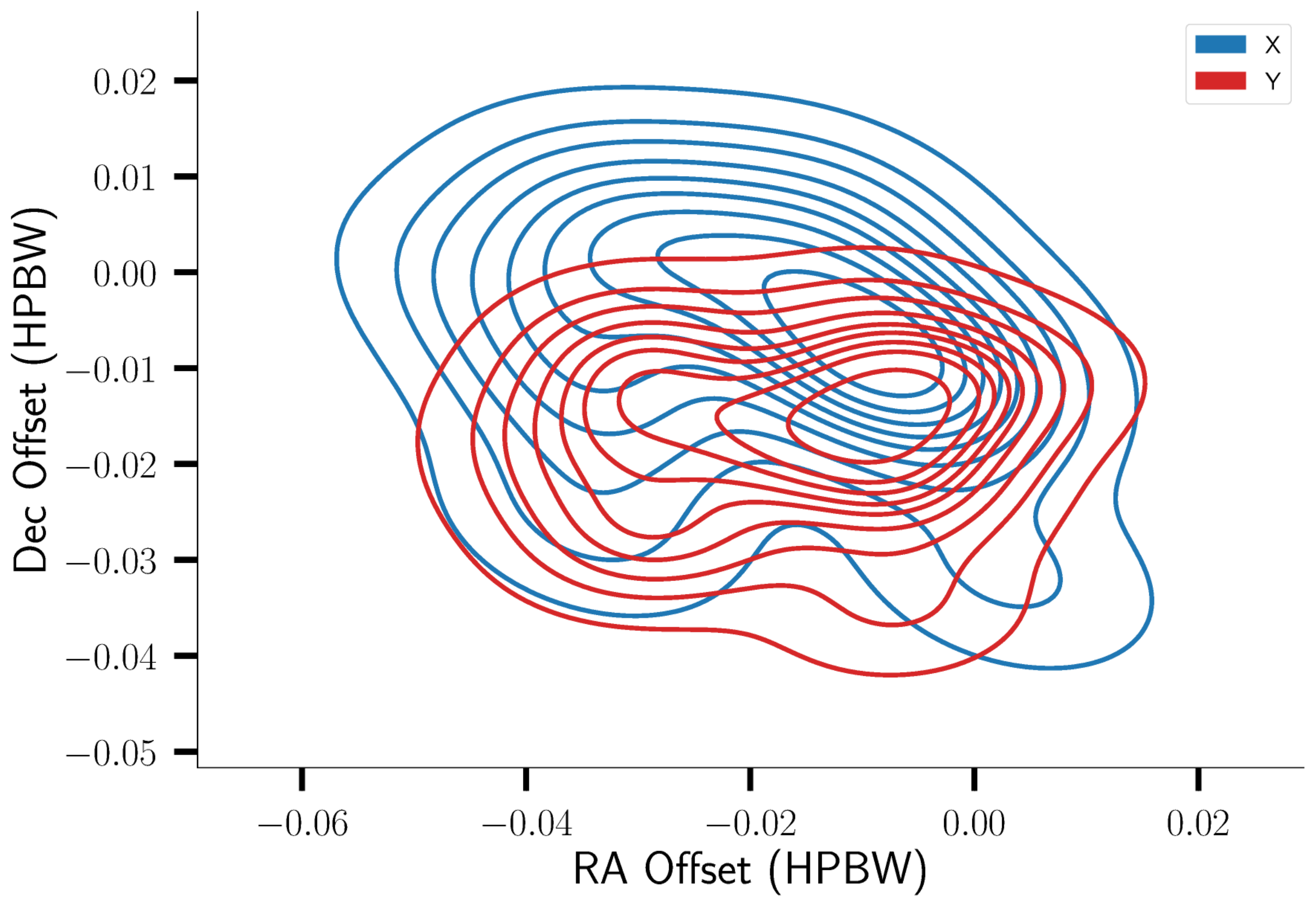}{0.5\textwidth}{(a) ALMA DA}
    \fig{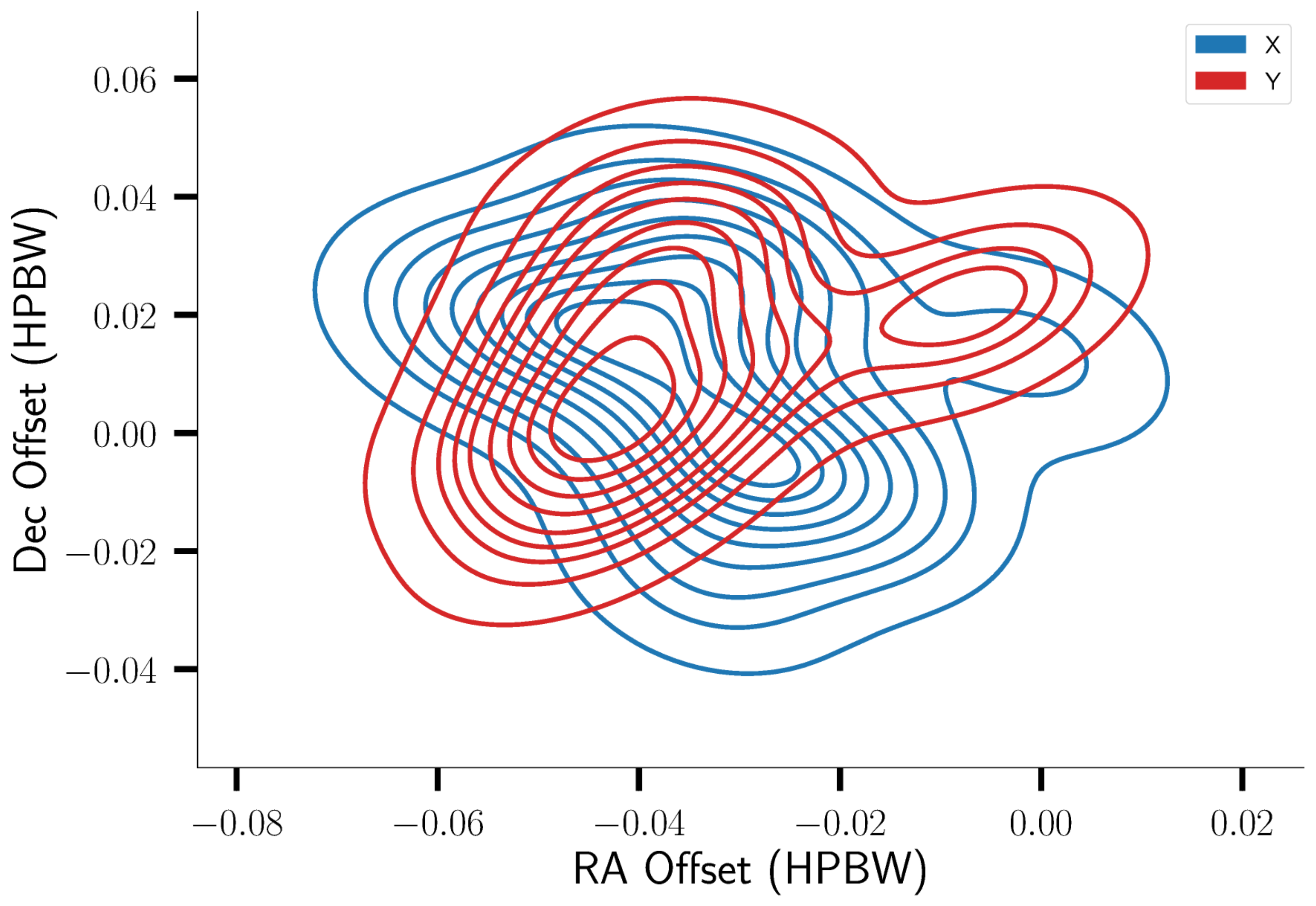}{0.5\textwidth}{(b) ALMA DV}
  }

  \caption{Contour density plots of the measured pointing offsets for VLA, MeerKAT, ALMA
          DA and DV.  In all cases, the two different colors correspond to the different
          correlation products.  VLA is the only antenna type with circular feeds, and it
          shows a clear separation between the pointing centres of the R \& L feeds. This
          separation is the well-known beam squint. The squint as a function of frequency
          is plotted in Fig.~\ref{fig:beam_squint_squash_freq}. MeerKAT and ALMA both have
          linear feeds and do not show a consistent separation between the two feeds.}

  \label{fig:pointing_offsets}
  \end{figure*}

\begin{figure*}
  \gridline{\fig{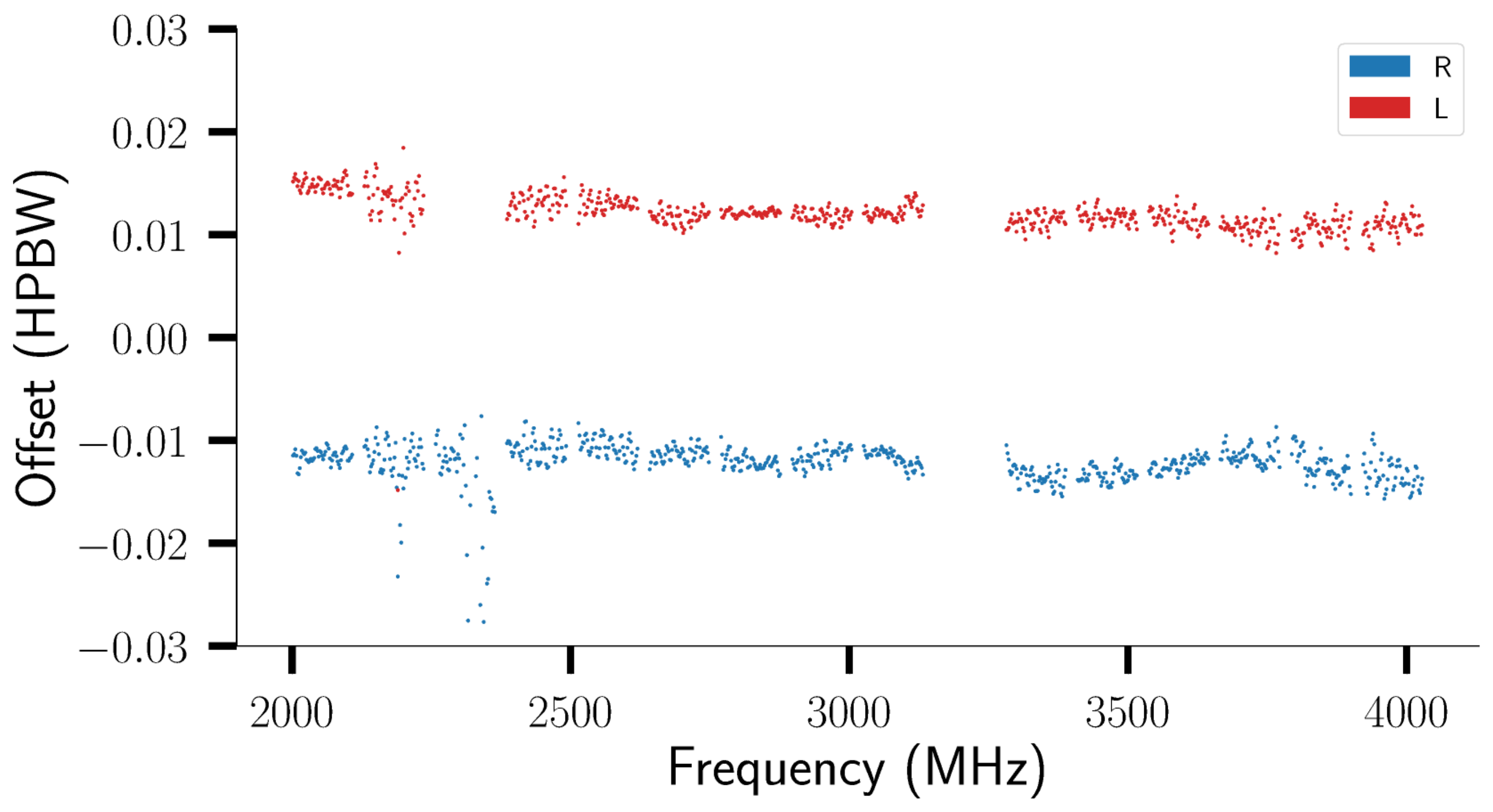}{0.5\textwidth}{(1a) EVLA Mean Offset}
  \fig{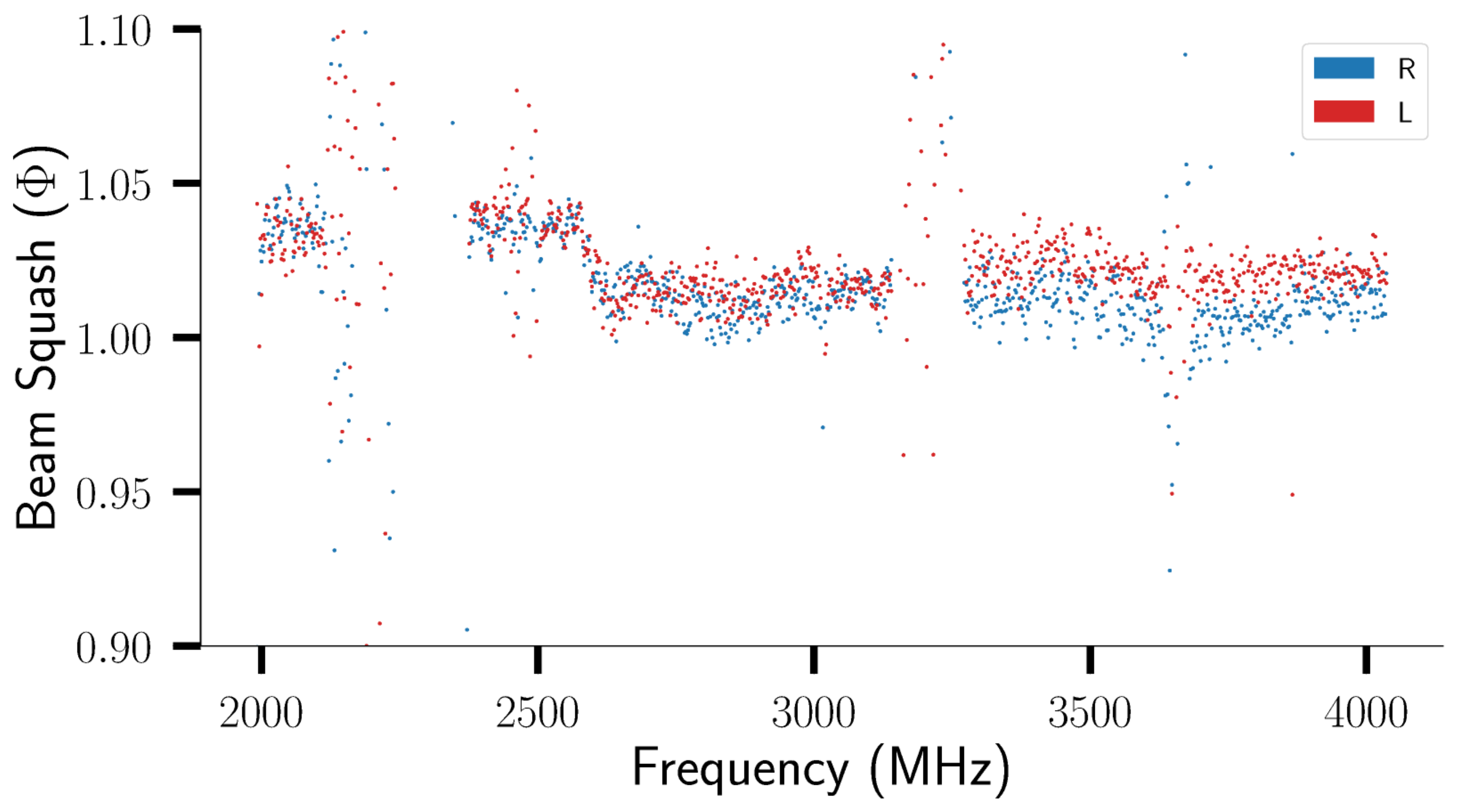}{0.5\textwidth}{(1b) EVLA Squash}}
   \gridline{\fig{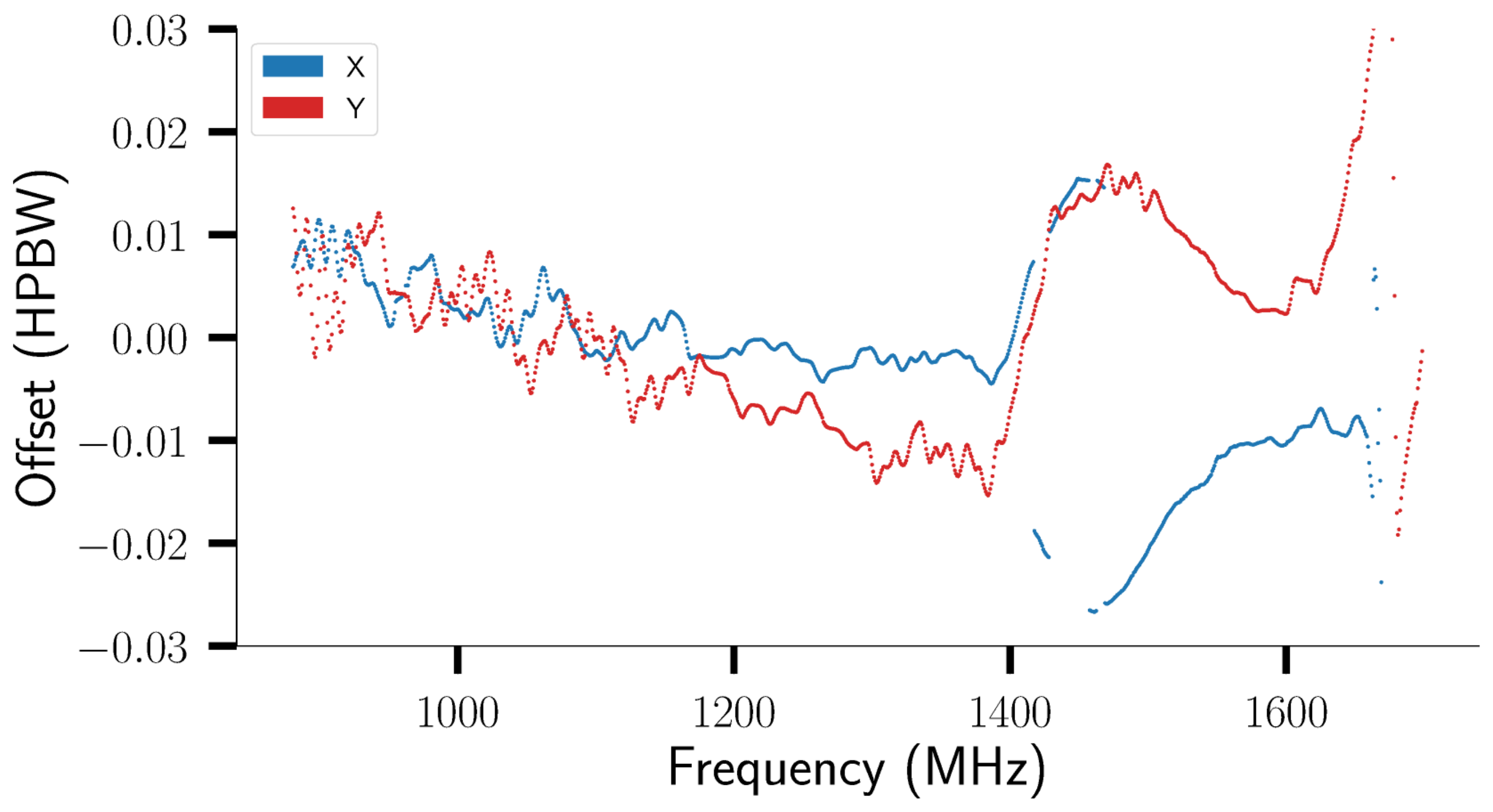}{0.5\textwidth}{(2a) MeerKAT Mean offset}
            \fig{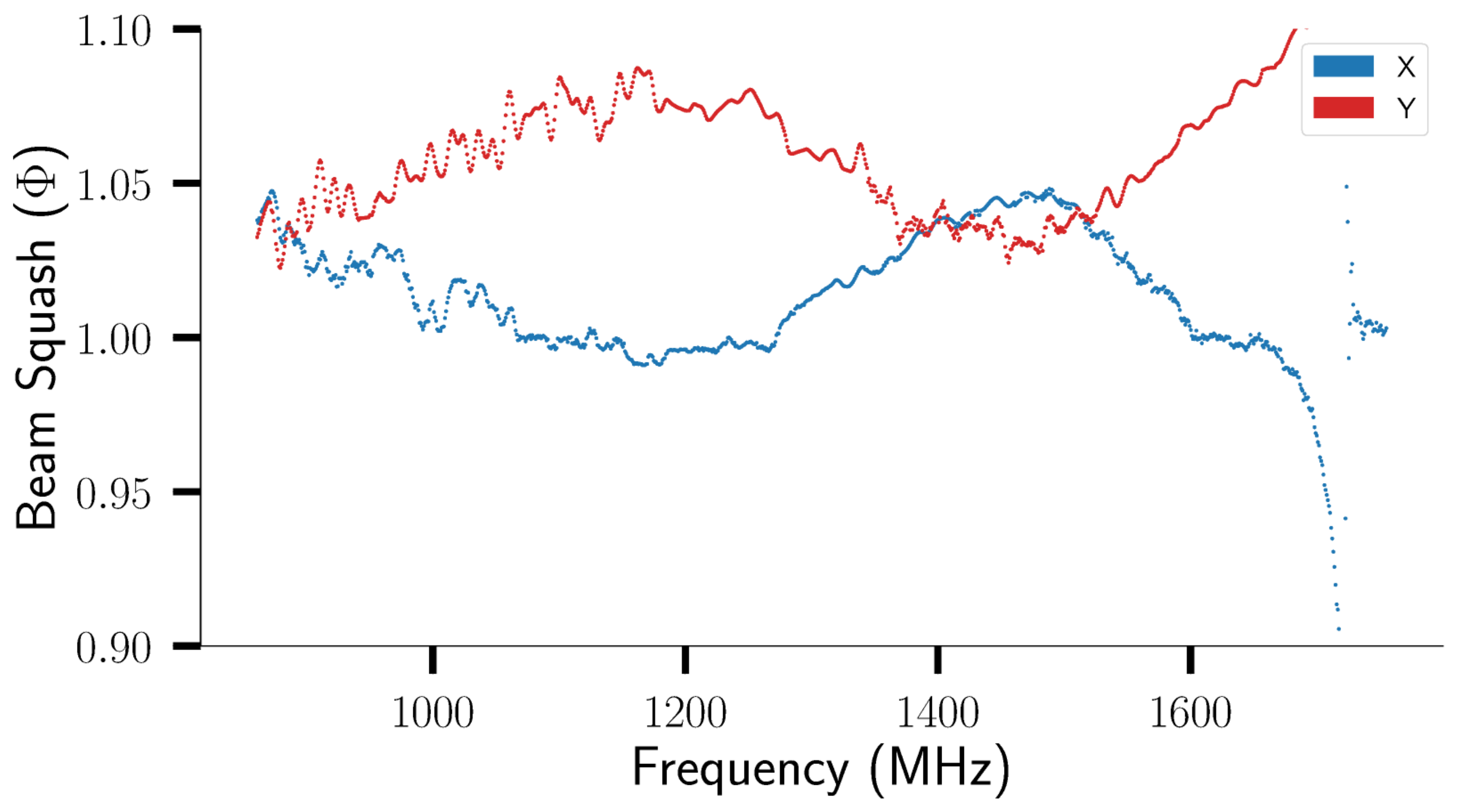}{0.5\textwidth}{(2b) MeerKAT Squash}}
          \caption{Plots of the mean pointing offset (left) and the beam squash
          (right) per correlation product as a function of frequency for EVLA and MeerKAT.
          These data show the frequency dependence of the same offsets plotted in
          Fig.~\ref{fig:pointing_offsets}.
          \textbf{(1)} VLA shows a clear separation between the R \& L
          feeds, which corresponds to $\sim 3\%$ of the beam width as a function of
          frequency. This is consistent with previous measurements of beam squint for
          VLA.  On the other hand, although the VLA PB shows some degree of
          ellipticity ($\Phi > 1$), the R and L beams are very similar across
          the band, and do not show a large difference in their beam squash.
          \textbf{(2)} MeerKAT has a more complex beam squint behaviour vs.
          frequency. Both the X and Y feeds tend toward a zero mean offset at
          higher frequencies, with a minimum at $\sim 1400$ MHz. The X and Y
          beams show a significantly different beam squash, indicating that the
          beams are preferentially elongated along orthogonal axes. Similarly to the squint,
          the beam squash is a minimum at $\sim 1400$ MHz.}
  \label{fig:beam_squint_squash_freq}
\end{figure*}

Fig.~\ref{fig:pointing_offsets} shows the contour density plots of the measured residual
pointing errors for EVLA, MeerKAT, ALMA DA \& DV antennas. The errors plotted
here were determined independently per antenna, SPW and channel, except for
ALMA, for which the band averaged values for a single SPW were used. In all
cases each colour corresponds to one of the two orthogonal feeds (blue = R \& X;
red = L \& Y respectively). The VLA shows a clear separation between the pointing
centres of the R \& L feeds, corresponding to the well known beam squint as demonstrated
in \citet{jagannathan2017directionii}. Both MeerKAT and ALMA show significant overlap
between the pointing centre of the feeds, as is expected from basic antenna optics for
linear and circular feeds.

Figure~\ref{fig:beam_squint_squash_freq} plots the frequency dependence of the
mean squint and the beam squash. The mean pointing offset for each correlation
is defined as

\begin{equation}
  d_{\text{offset}}(\nu) = \frac{1}{N_{\text{ant}}}\sum_{i}\frac{\Delta \alpha_{i}(\nu) + \Delta \delta_{i}(\nu)}{2}
  \end{equation}

\noindent
where $\Delta \alpha_{i}(\nu)$ and $\Delta \delta_{i}(\nu)$ are the right
ascension and declination offsets respectively as a function of frequency. The
beam squash is defined as

\begin{equation}
 \Phi = \frac{\overline{l_{\text{major}}}}{\overline{l_{\text{minor}}}}
\end{equation}

where $\overline{l_{\text{major}}}$ and $\overline{l_{\text{minor}}}$ are
the major and minor axes of the ellipse fitted to the beam of the antenna
averaged response in feed basis, measured along the position angle of the
ellipse. The plot only shows the frequency dependence for VLA and MeerKAT since
they have a large fractional bandwidth.

The VLA shows a clear separation between the pointing centres
of the R and L feeds (Fig.~\ref{fig:beam_squint_squash_freq}(1a)). The
separation is $\sim 3\%$ of the HPBW as a function of frequency, which is
consistent with previous measurements of the beam squint.  MeerKAT however shows
a more complex beam squint behaviour across the band. Both X \& Y feeds are
biased toward positive offsets at lower frequencies, and approach a minimum at
$\sim 1400$ MHz. The beam squint also shows a quasi-oscillatory behaviour. The
periodicity of the oscillation does not correspond to the standing wave
frequency, and the cause of the quasi-periodicity of the pointing is not
immediately obvious. In terms of beam squash, VLA shows consistent behaviour
across the band. There is a small amount of beam ellipticity
($\Phi = 1.01 - 1.03$) increasing slightly toward the higher frequencies.
MeerKAT beam squash behaviour is analogous to the beam squint, the two feeds
showing ellipticity along orthogonal directions, with a minimum at $\sim 1400$
MHz.

\subsection{Direction Dependent Polarization Leakage}

\begin{figure*}
  \gridline{\fig{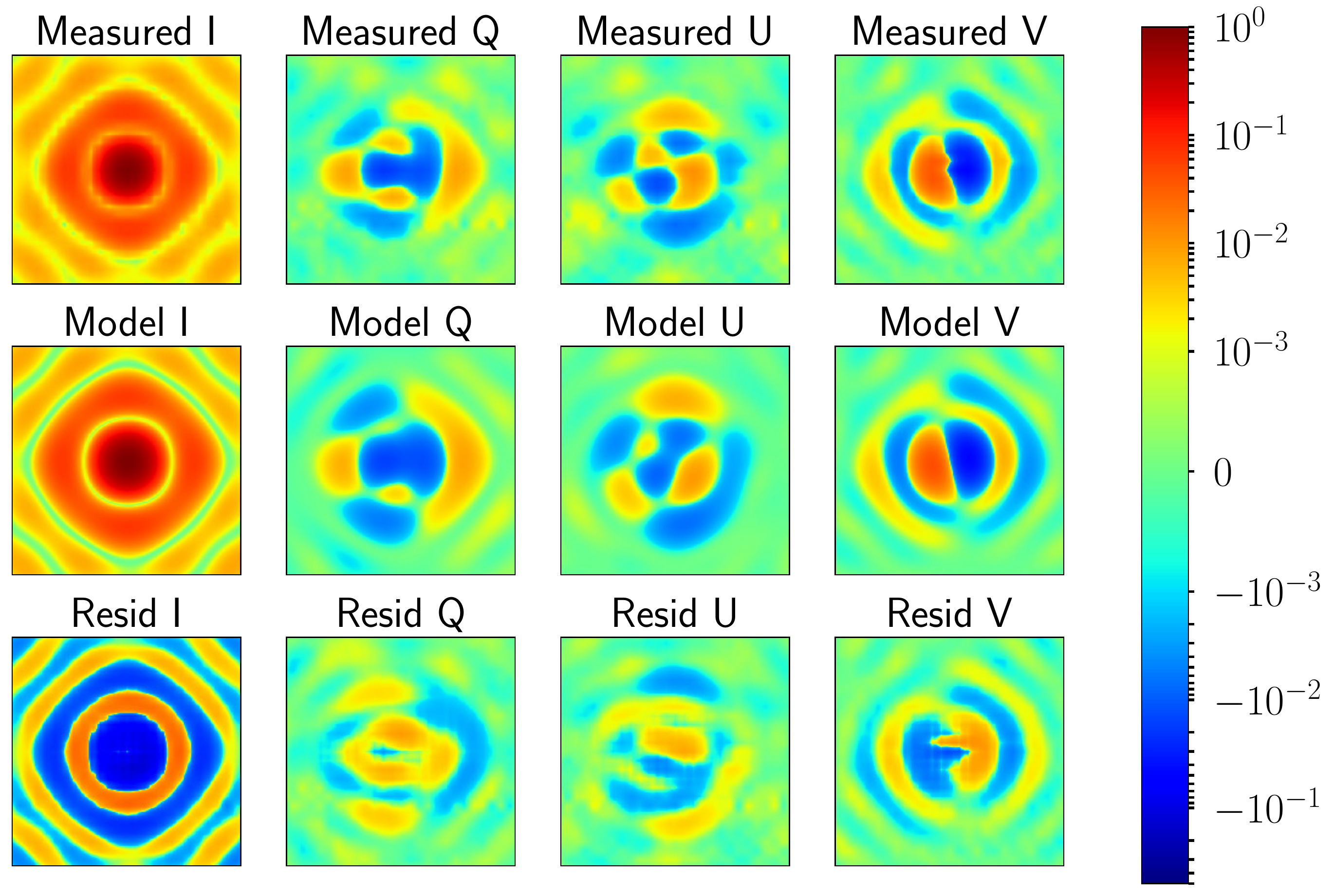}{0.5\textwidth}{(a) EVLA}
            \fig{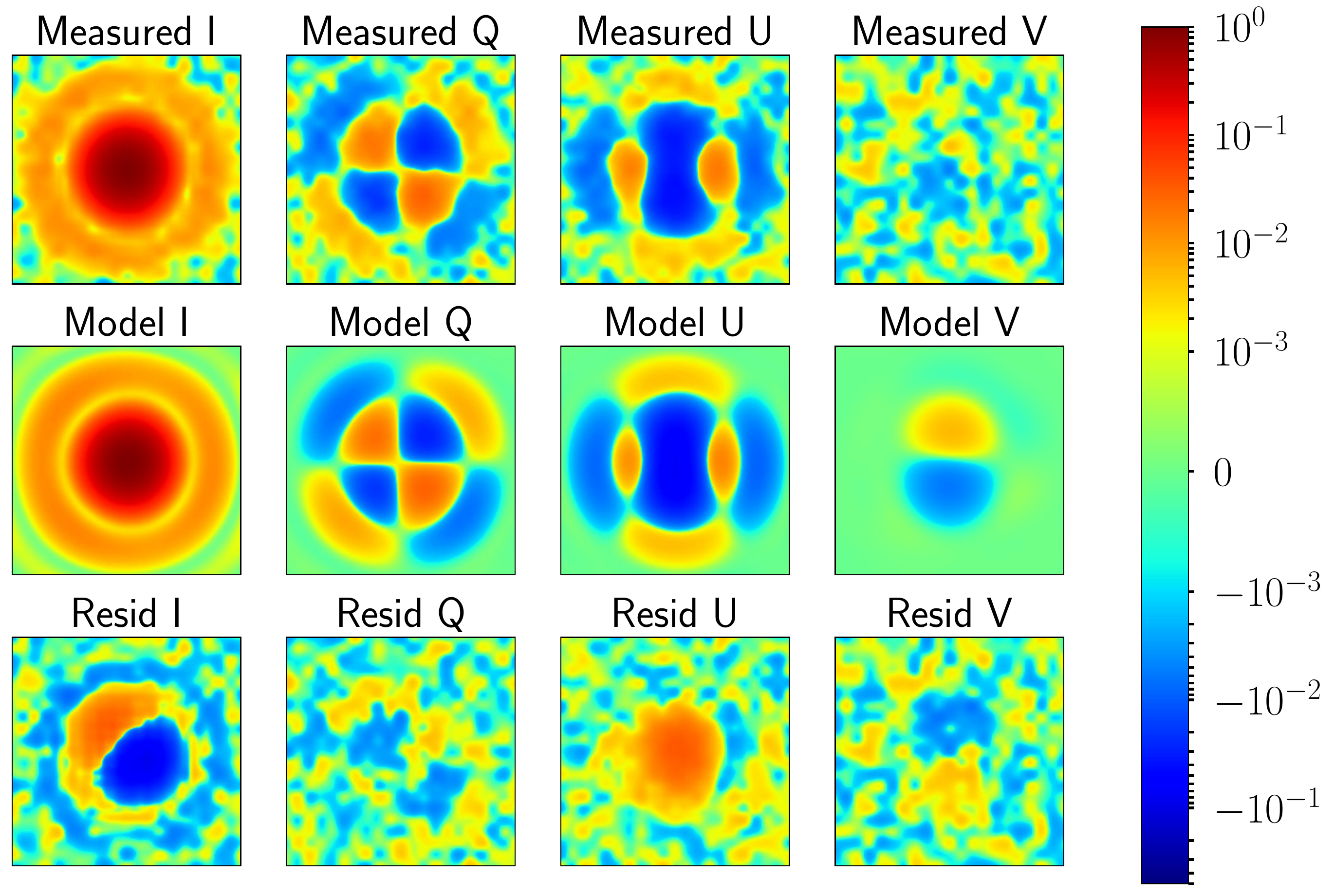}{0.5\textwidth}{(b) ALMA DA}}
  \caption{The full Mueller residuals for VLA (left) and ALMA (right). In each
          figure, from top to bottom are the measured Stokes beams (total power), the model Stokes beams
          and the residual. There is good agreement between the morphology of
          the measured and model Stokes beams. For ALMA, the Stokes V
          measurement has very low SNR since it is the difference between the cross-hand
          feeds. We note that this is a problem only in the total power
          measurements, the Jones beams have better SNR.}
  \label{fig:mueller_resids}
\end{figure*}

During the process of antenna holography, we obtain two independent measurements
of the antenna beams. The first are the Jones beams, as we discussed
in previous sections. The second are the total power beams (\textit{cf.}
Eq.~\ref{eq:holo} and Eq.~\ref{eq:total_power_holo}). Constructing the Stokes
beams from the total power beams is a much more straightforward process, and is
identical to constructing the Stokes products from the correlated visibilities
themselves. By comparing the full Stokes beams derived from Zernike  models to the total power
beams, we are able to verify both the efficacy of the Jones modeling against an
independent measurement as well as the sign conventions of the Jones to Mueller
unitary conversion matrix $\matr{S}$ \citep{hamaker1996understanding}. The
conversion from Jones to Mueller via the unitary matrix S is given by
\begin{equation}
  \matr{M}_{\text{stokes}} = \matr{S}^{\dagger}\left[ \matr{J}_{i}\otimes \matr{J}_{j}\right]_{\text{feed}}\matr{S}
\end{equation}

\noindent
where $\matr{S}$ is a $4\times 4$ transformation matrix, and $\matr{J}_{i}$ and $\matr{J}_{j}$
are the $2\times 2$ Jones matrices for antennas $i$ and $j$ respectively.

Fig.~\ref{fig:mueller_resids} shows the measured, model, and residual VLA S Band
and ALMA Band 3 Stokes products. The top row shows the measured Stokes products
from the total power beams.  The second row shows the equivalent Zernike models
of the beams, and the bottom row shows the  difference between the two. The
residuals in Stokes I are at the level of \textit{i.e.,} a $\sim 1\% $
fractional error on the beam model.  The morphology for the Stokes-Q and -U
beams are generally in agreement.  We note that the power beams are different from
the Jones beams in two ways - \textit{(i)} the baseline pairs used to construct
the two sets of beams are different (and non-overlapping), and \textit{(ii)} the
effect of pointing errors will affect the two beams slightly differently. The
comparison with the power beams or the Mueller matrix does not reflect the
efficacy of the fit itself, but rather demonstrates internal consistency of the
Jones matrices and the transformation matrix required to go from feed to stokes
basis, namely the $\matr{S}$ matrix.

As an additional consistency check, we compare the Stokes I PB generated from AW
project to the Stokes I PB from holography. AW project calculates the PB
via a Fourier transform of the gridded, weighted baseline AIP derived during
imaging which is reflective of the true PB of the measurement.

\begin{figure}
  \gridline{\fig{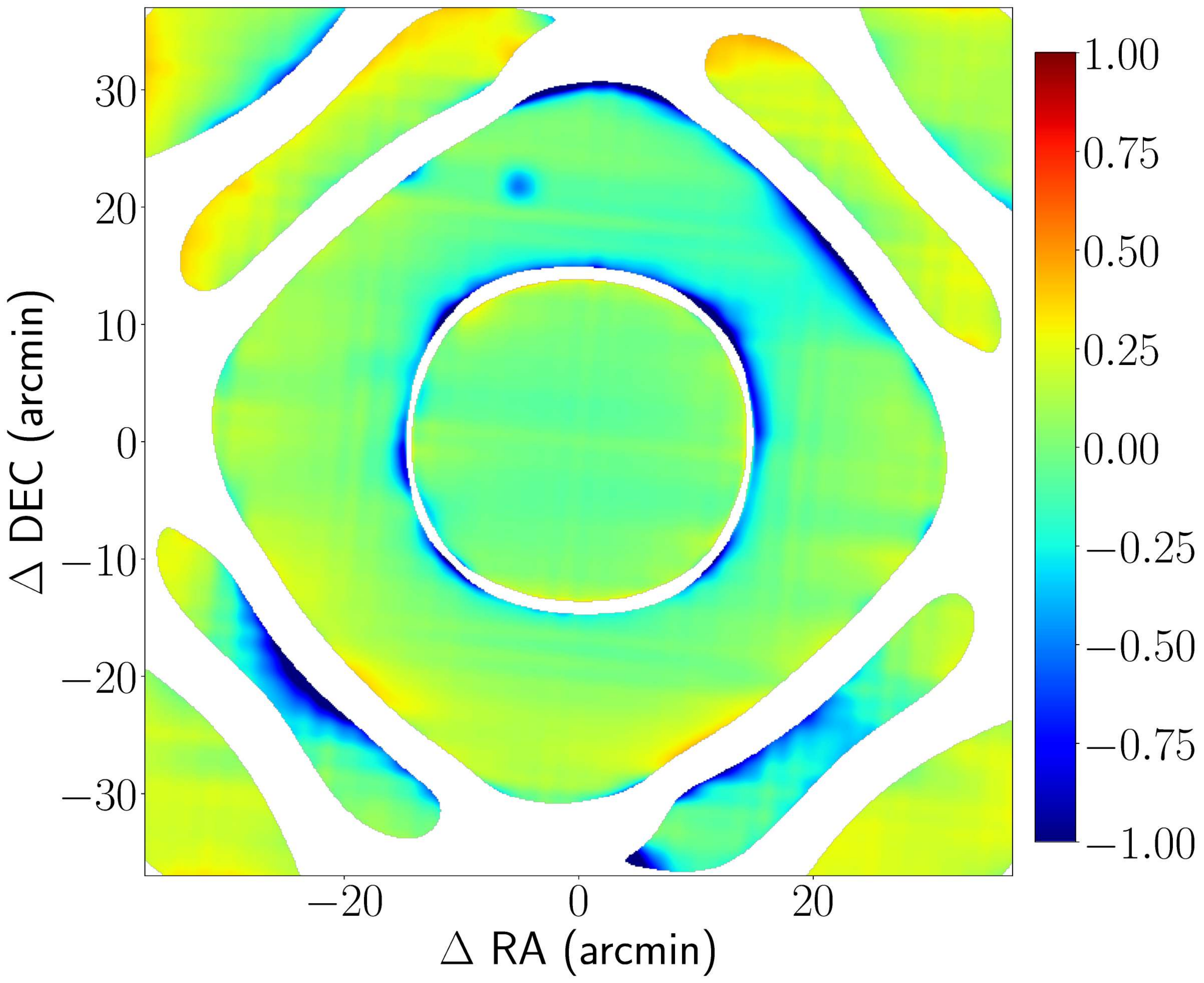}{0.5\columnwidth}{(a) EVLA}
            \fig{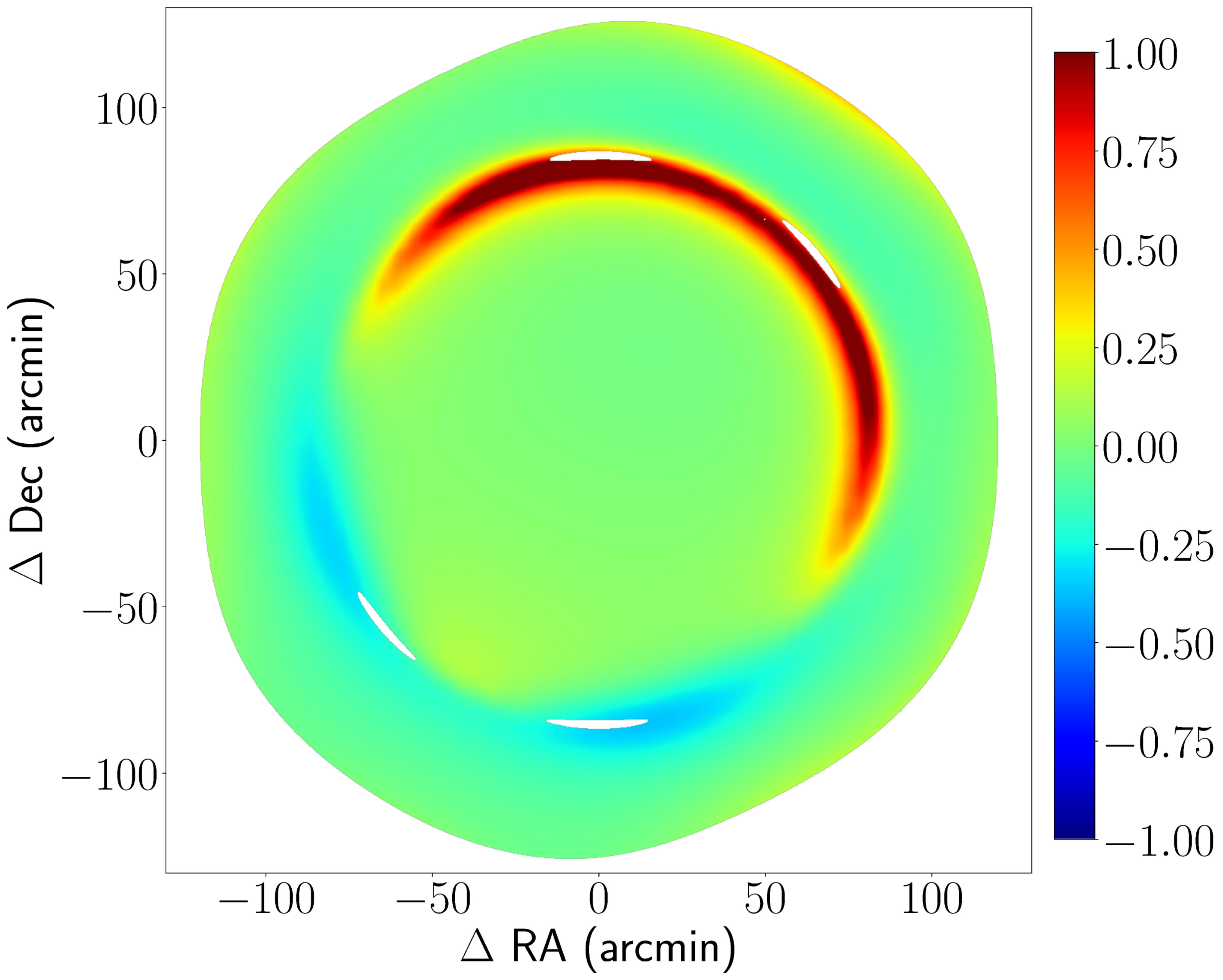}{0.5\columnwidth}{(b) MeerKAT}}
  \caption{Fractional residual between the Stokes I PB produced by the AW
    projection framework and the Stokes I PB generated from measured holography
    for VLA at 2.9 GHz (left) and MeerKAT at 1.25 GHz (right). In both cases
    the fractional error is $\leq 2\%$ within the main lobe, rising to $\leq 10\%$
    in the first sidelobe. The fractional error near the nulls will naturally
    tend to be very large.}
  \label{fig:fractional_pb_compare}
\end{figure}

Fig.~\ref{fig:fractional_pb_compare} shows the fractional residuals between the
Stokes I beam produced by the \texttt{awproject} gridder in CASA and the Stokes I beam
derived from the holography for VLA at 2.8 GHz and MeerKAT at 1.25 GHz.
This figure is representative of other SPWs across the band for both telescopes.
The fractional error is large toward the nulls, which is expected. The AW
projection code cannot as yet perform full Mueller corrections and hence we
restrict ourselves to comparing the Stokes I beams. From these plots it is clear
that our modeling has captured the PB response well, both in the main lobe and
the first (few) sidelobe(s). The details of how these coefficients are included
in CASA are outlined in Appendix \ref{appendix:B}.
The implementation of a full-Mueller, wideband AW projection algorithm is
currently being tested, the results of which will be described in a forthcoming
paper (Jagannathan et al., \textit{in prep}).

\subsection{Modeling the aperture across frequency}
\label{sec:aperture_eta}

\begin{figure}
  \plotone{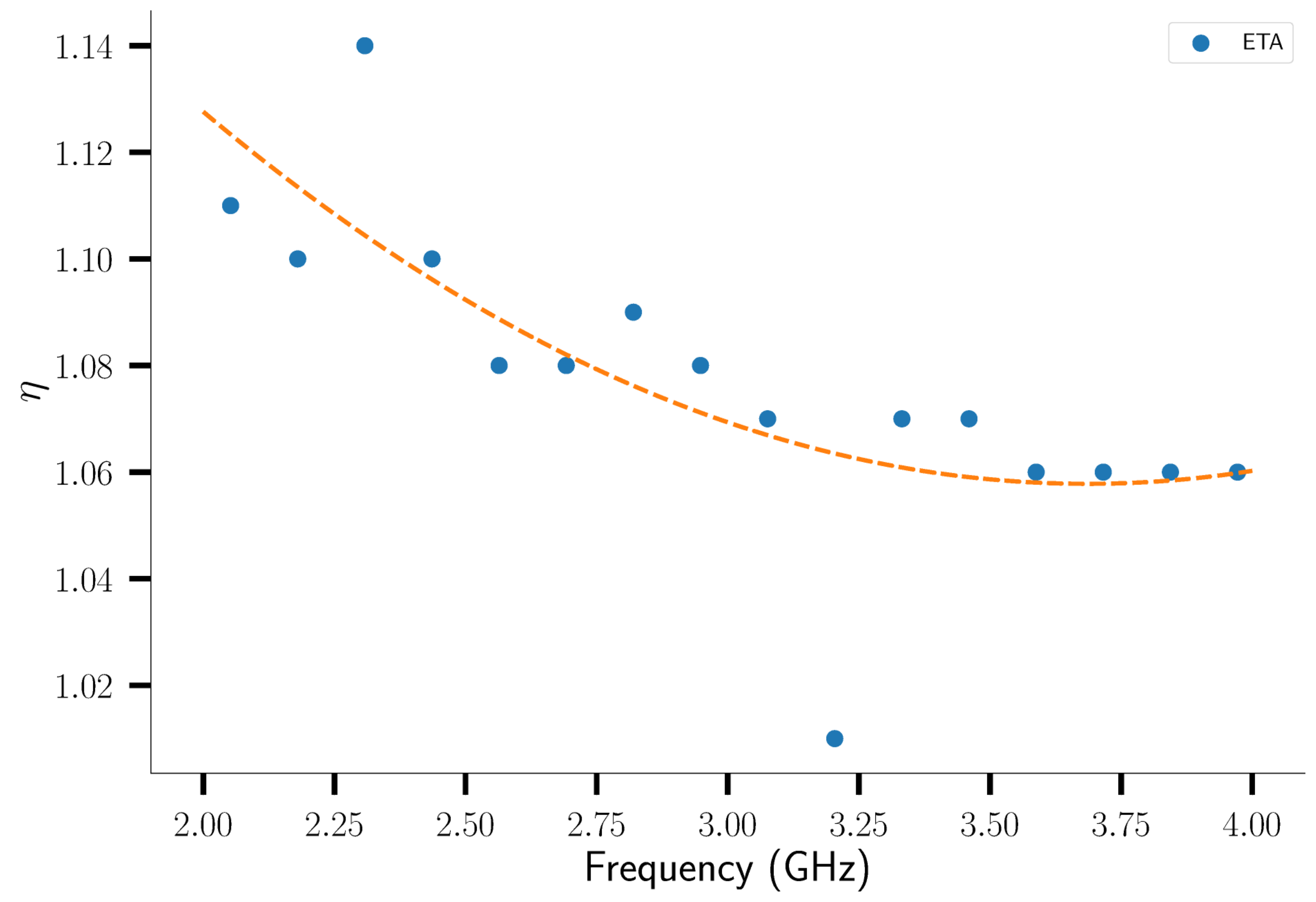}
  \caption{Plot of $\eta_{A}$ scaling factor as a function of frequency for VLA
  S Band.  The dashed line is the polynomial fit to the data. We find that
  $\eta_{A}$ reduces as a function of frequency, analogous to increasing aperture
  efficiency, and flattens out at the highest frequencies. The sudden spike at
  $\sim$ 2.25 GHz and dip at $\sim$ 3.25 GHz correspond to SPWs corrupted by RFI,
  yielding unreliable antenna holography measurements.}
  \label{fig:eta_ap_efficiency}
\end{figure}

In order to obtain PB sizes accurate to $< 1\%$ we need to account for changes
in aperture efficiency across the band. Aperture efficiency changes the
effective diameter of the dish, and we introduce a scaling factor $\eta_{A}$ to
account for this as follows:

\begin{align}
  D_{A}^{\prime}(\nu) = \eta_{A}(\nu) D_{A}
\end{align}

where $D_{A}$ is the nominal antenna diameter (25m in the case of VLA) and
$D_{A}^{\prime}$ is the effective antenna diameter.
Fig.~\ref{fig:eta_ap_efficiency} plots the behaviour of $\eta_{A}$ as a function
of frequency for VLA S-Band observations.

We derive this scaling factor in the following manner : \textit{(i)} $\eta_{A}$
is introduced in the code as a specifiable (free) parameter \textit{(ii)} For a
given value of $\eta_{A}$ a primary beam image is generated (at a given
frequency) using the \texttt{awproject} gridder in CASA, and the residuals are
computed against the PB generated via the \texttt{standard} gridder. The value
of $\eta_{A}$ that minimizes these residuals is stored. We found that
a ``brute-force'' minimization method was sufficient,
for generating a PB to the accuracy we required. To that end at every $\eta_{A}$
starting from 0.7 through 1.3 in steps of 0.01 the minimization was carried out.
Further reducing the step size to 0.001 yielded no improvement in the values of $\eta_{A}$.

\subsection {Efficacy}

\begin{figure}
  \plotone{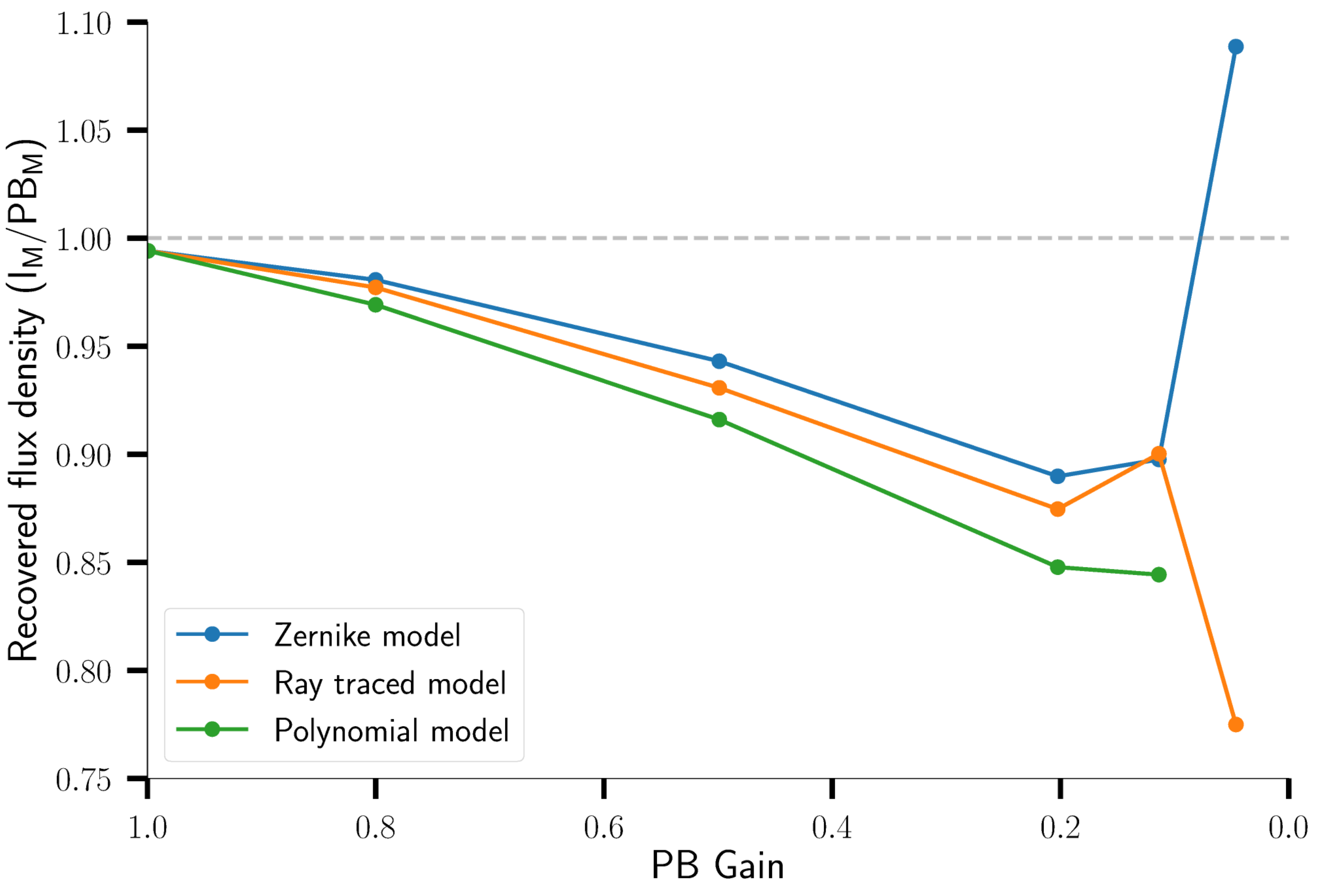}
  \caption{Plot of the recovered flux between the Zernike model, the ray traced
    model and the polynomial model of the EVLA PB.}
  \label{fig:recovered_flux_compare}
\end{figure}

To show the performance of the Zernike modelled apertures in imaging we carried
out a simulation. We generated a sky model of 6 point sources, each of flux
density 1 Jy which were then attenuated by the average measured holographic
antenna PB, using a central frequency of 2948 MHz and 64MHz bandwidth. The
simulated measurement set is generated following
\citet{jagannathan2017direction}.

The sources were placed at PB gain locations of 1.0, 0.8, 0.5, 0.2, 0.1, and
0.05.  The first five sources were placed in the main-lobe and the last source
at the peak of the first side-lobe.  The resulting measurement set was then
imaged using the \texttt{awproject} and \texttt{standard} gridding algorithms
in CASA and the resulting images were primary beam corrected to recover
the true flux densities \textit{i.e.,} $\vect{I}_{M}/\vect{PB}_{M}$ where
$\vect{I}_{M}$ is the recovered sky model and $\vect{PB}_{M}$ the antenna
primary beam produced by the imaging algorithm during imaging.

Fig.~\ref{fig:recovered_flux_compare} shows the recovered flux densities of the
sources as a function of the PB gain level. The dashed line in grey represents
the true flux density of the predicted sources. If the recovery is exact all
points would lie along that line. The blue line represents the new Zernike
aperture models used in AW-Projection while the orange line represents
ray-traced aperture illumination models currently in use. The green line plots
the polynomial model of the antenna main-lobe utilized by the default gridding
algorithm (\textit{i.e.,} the \texttt{standard} gridder) within the CASA
\texttt{tclean} framework. The figure demonstrates that the Zernike polynomial
model more accurately represents the antenna holography and consequently does a
better job of retrieving the flux density of the source across the main-lobe and
out to the first side-lobe where the error in the retrieved flux density is
~$10\%$. The ray-traced AIP performance within the main-lobe is consistently
lower by $2\%$ across the main-lobe but underestimates the flux density by
nearly $~23\%$ at the first side-lobe. The improvements in modeling the
sidelobe, as well as the ability to generate full-Mueller PB models is the
primary advantage of the method outlined in this paper.  The level of flux
density recovery makes the A-to-Z solver a more versatile and effective method
than both ray tracing and using axis symmetric 2D polynomial as demonstrated
here.

\section{Conclusion}
\label{sec:conclusion}

We have demonstrated the A-to-Z solver methodology which we use to model the
full Jones response of an antenna using Zernike polynomials. We have further
demonstrated that this approach results in widefield, wideband, full Mueller PB
models that are accurate to the first sidelobe. By using the measured AIP, we
rely on direct measurements of the optical properties to inform the modeling,
which is necessary in capturing the polarization leakage behaviour. This makes
it easy to generalize our method to any arbitrary interferometer with
holographic measurements, without needing to rely on setting up complex
simulations that typically require a large time and compute investment. We have
demonstrated the generality of our approach by modeling the VLA, ALMA and
MeerKAT telescopes which have a variety of different feed polarization, dish,
and frequency configurations. The only limitation to extending this method to
other telescopes and facilities is the availability of high quality
interferometric observations.

We also demonstrate the efficacy of the Zernike polynomials in modeling optical
properties of the dish and beam, such as the standing wave due to the second
reflection between the antenna feed and secondary reflector. These effects show
up in different Zernike terms, that typically correspond to actual optical
aberrations such as the tip, tilt, defocus and astigmatism. The broadband beam
squint and beam squash behaviour are also well modeled, and our measurements are
consistent with existing previous estimates.  The primary beam models described
in this paper are generally accurate to one to two sidelobes, which is
relevant for widefield imaging and deep mosaicing experiments.

We have implemented the A-to-Z modeling functionality in a Python package \citep{Sekhar_ZCPB_2021}
\footnote{https://gitlab.nrao.edu/pjaganna/zcpb} that (at the time of writing)
has been verified to work on VLA L- and S-Band, MeerKAT L-Band and ALMA Band 3
data. These models can then be used to generate full Stokes beam models
(\textit{i.e.,} the first row of the Mueller matrix), the functionality for
which has been implemented in a separate Python package \citep{Sekhar_Plumber_2021}
\footnote{https://github.com/Kitchi/plumber} in order to perform image
plane leakage corrections. These coefficients have also been included in (at the
time of writing) a development branch of CASA that uses the Zernike
models within the A-Projection framework. As mentioned earlier, this branch
currently only corrects for Mueller-diagonal terms and the full Mueller
corrections are underway. We will present the details of both the aperture and
image plane widefield polarization leakage corrections in a forthcoming paper in
this series (Jagannathan et al., \textit{in prep}).

\appendix

\section{Appendix A : FM Models}
\label{appendix:A}

\begin{figure*}
  \gridline{\fig{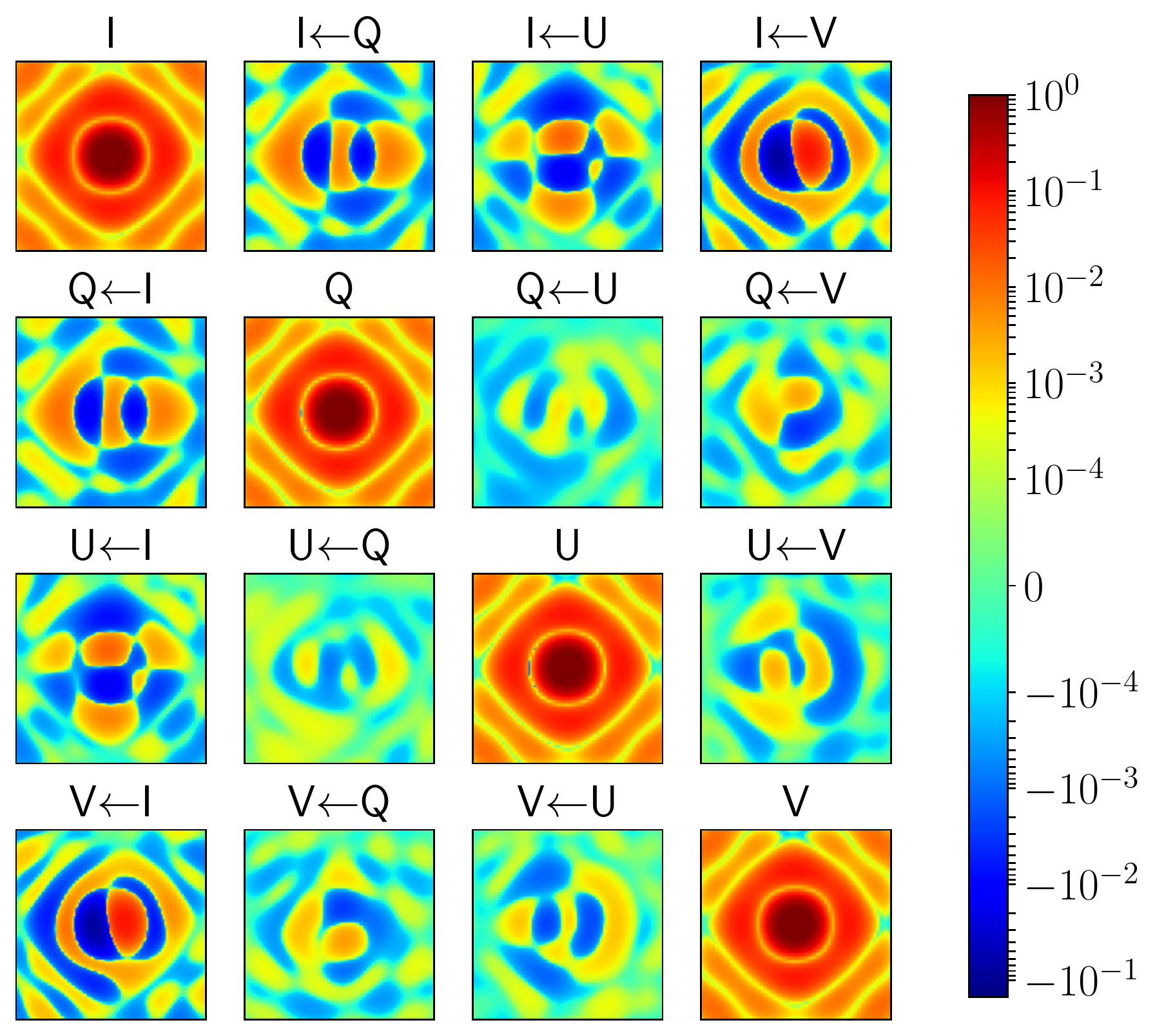}{0.5\textwidth}{(a) EVLA}
            \fig{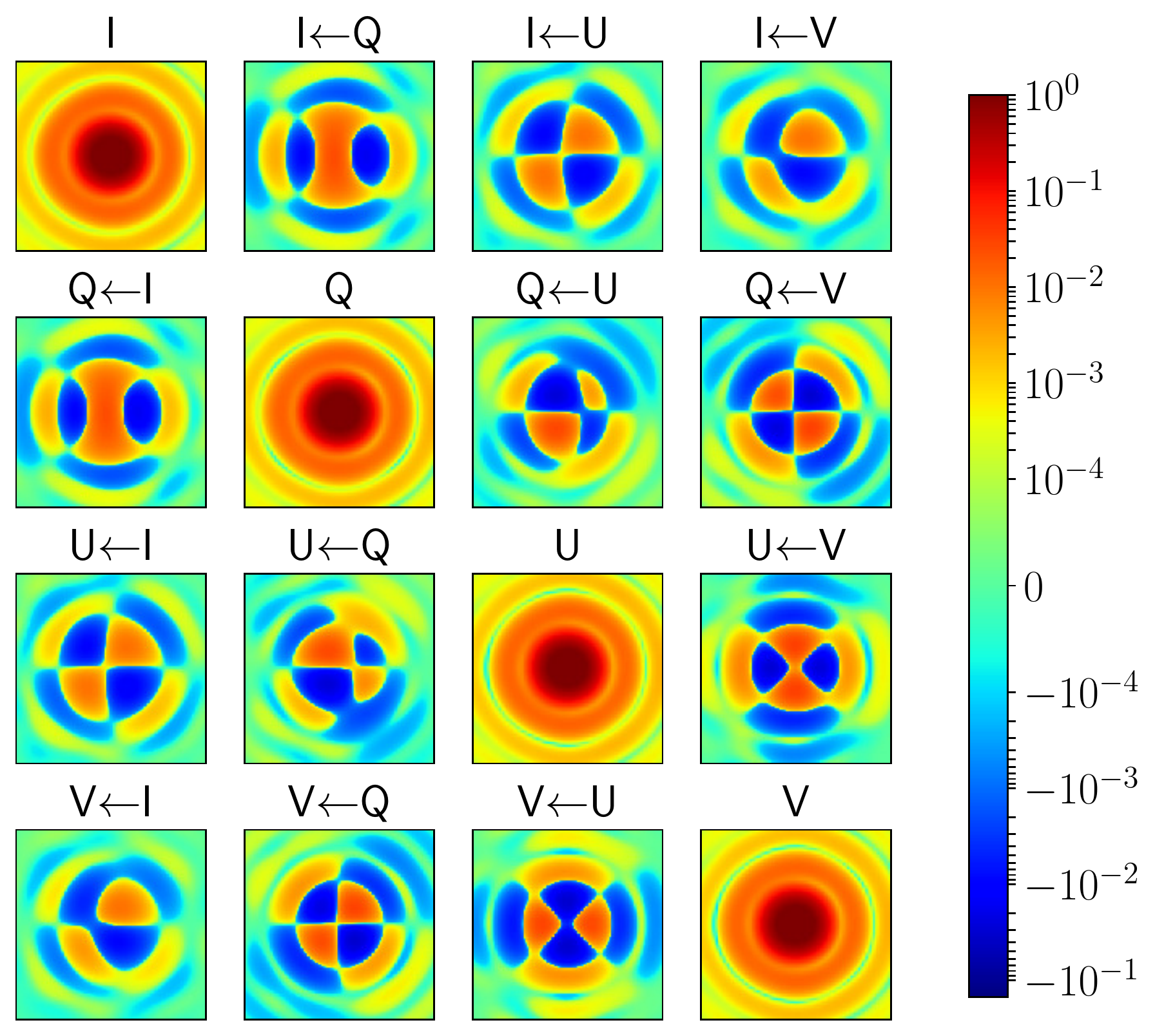}{0.5\textwidth}{(b) MeerKAT}}
  \gridline{\fig{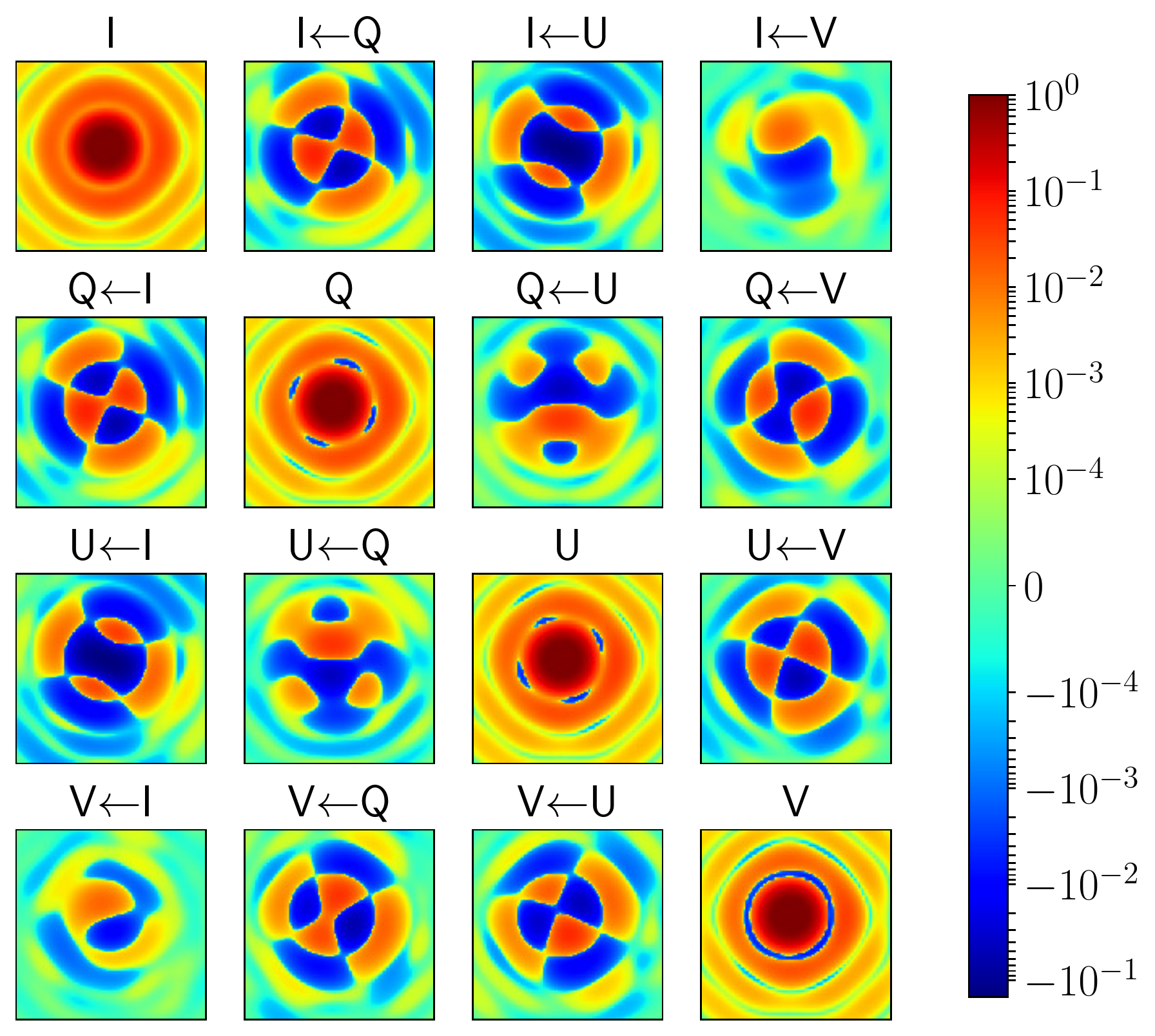}{0.5\textwidth}{(c) ALMA DV}
            \fig{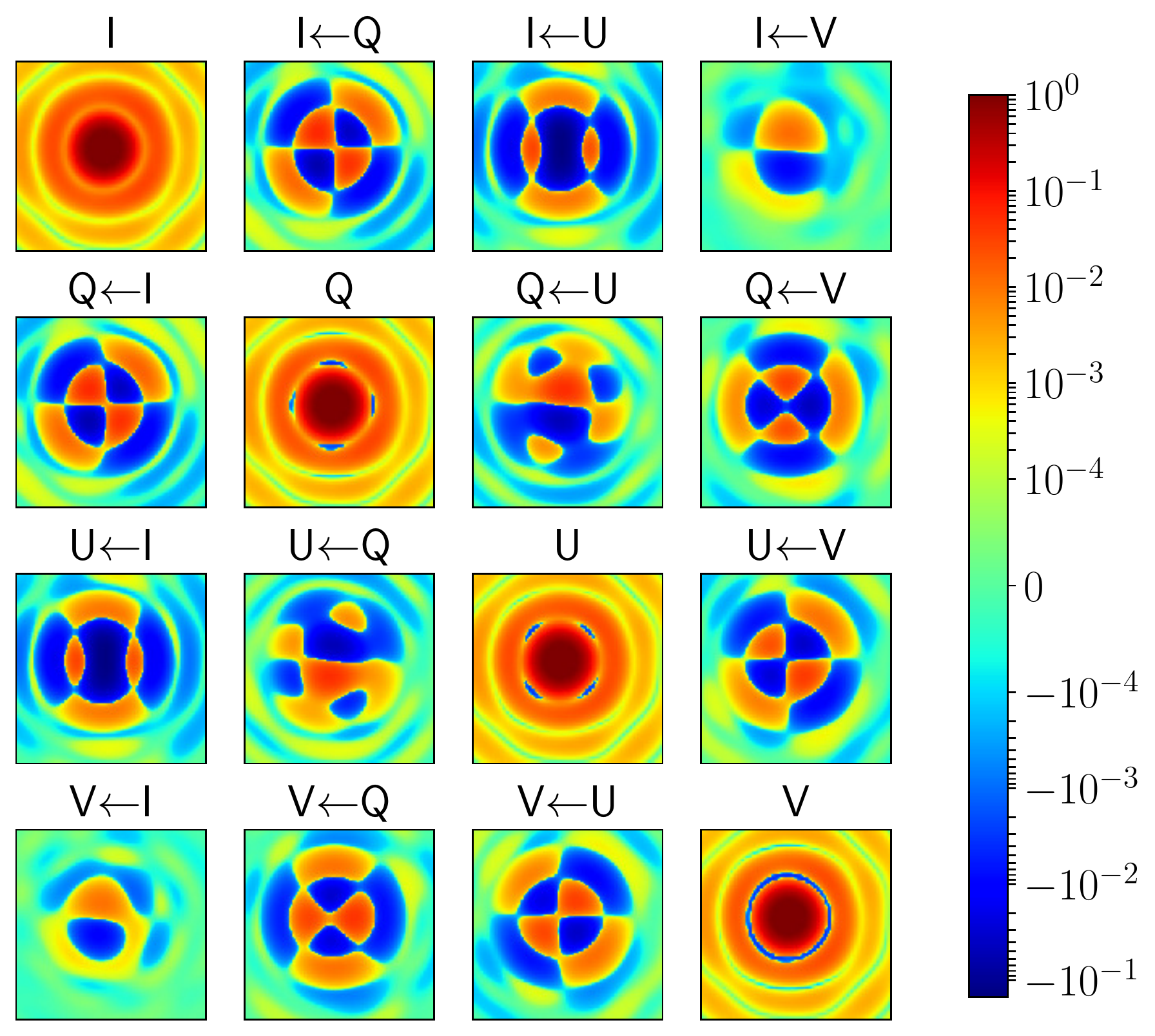}{0.5\textwidth}{(d) ALMA DA}}
          \caption{The direction dependent Mueller terms generated from the Zernike models for (a) VLA S Band (2.5 GHz) (b) MeerKAT L Band (1.2 GHz) (c) ALMA DV antenna Band 3 (108 GHz) (d) ALMA DA antenna Band 3 (108 GHz). }
  \label{fig:zernike_muellers}
  \end{figure*}

Figure \ref{fig:zernike_muellers} plots the Full Mueller models for all the
telescopes discussed in this paper. These models were obtained by constructing
the Mueller elements from the Jones models.

\section{Implementation of Zernike Coefficients in \texttt{tclean}}
\label{appendix:B}

The coefficients derived in the manner described in this
Sec.~\ref{sec:aperture_fitting} are included in the \texttt{awproject} gridder
in CASA by listing them in a CSV file. This format allows for the
specification of Zernike coefficients as a function of frequency, polarization
and antenna type (when necessary).  Each band of each telescope will be
specified in a different CSV file that can be passed in to the code. Listing
\ref{lst:csv_file} shows an extract of such a CSV file. This is an extendable
format that can be modified to accept heterogeneous arrays (partially or fully)
by adding a further index to track the antenna type.

\begin{lstlisting}[caption={Extract from the CSV file specifying the cofficients for VLA S Band beams. The columns are Stokes, frequency (in MHz), the Zernike index, the real and imaginary coefficients, and the aperture efficiency factor.  The Stokes is listed in CASA readable format, using the numbers 5-8 to denote circular feeds and 9-12 for linear.}, basicstyle=\ttfamily\small, label=lst:csv_file]

  #stokes,freq,ind,real,imag,eta
  5,2052,0,308.13023030,-0.06968780,1.11
  6,2052,0,0.14375480,-0.07500360,1.11
  7,2052,0,0.11308890,0.24478490,1.11
  8,2052,0,301.83666030,-0.00622780,1.11

\end{lstlisting}Since Zernike polynomials are analytically well defined, the
aperture models can be exactly evaluated at the specified UV locations during
gridding, rather than using a cached pre-computed value from an over-sampled grid.

\begin{acknowledgements}

The National Radio Astronomy Observatory is a facility of the National Science
Foundation operated under cooperative agreement by Associated Universities, Inc.
The MeerKAT telescope is operated by the South African Radio Astronomy
Observatory, which is a facility of the National Research Foundation, an agency
of the Department of Science and Innovation.  IDIA is a partnership of the
University of Cape Town, the University of Pretoria and the University of the
Western Cape.  We acknowledge the use of the ilifu cloud computing facility -
www.ilifu.ac.za, a partnership between the University of Cape Town, the
University of the Western Cape, the University of Stellenbosch, Sol Plaatje
University, the Cape Peninsula University of Technology and the South African
Radio Astronomy Observatory. The ilifu facility is supported by contributions
from IDIA, the Computational Biology division at UCT and the Data Intensive
Research Initiative of South Africa (DIRISA).

This paper makes use of the following ALMA data:
$uid\_\_\_A002\_Xd2d9a0\_X8dd3$, $uid\_\_\_A002\_Xd2d9a0\_X92ae$,
$uid\_\_\_A002\_Xd2d9a0\_X991e$, $uid\_\_\_A002\_Xd2d9a0\_X8f4c$,
$uid\_\_\_A002\_Xd2d9a0\_X9427$, $uid\_\_\_A002\_Xd2d9a0\_X9b2e$.
ALMA is a partnership of ESO (representing its member states), NSF (USA) and
NINS (Japan), together with NRC (Canada), MOST and ASIAA (Taiwan), and KASI
(Republic of Korea), in cooperation with the Republic of Chile. The Joint ALMA
Observatory is operated by ESO, AUI/NRAO and NAOJ.

SS acknowledges financial support from the Inter-University Institute for Data
Intensive Astronomy (IDIA) and the National Radio Astronomy Observatory (NRAO).
IDIA is a partnership of the University of Cape Town, the University of Pretoria
and the University of the Western Cape.  PJ would like to thank Bruce Veidt and
Tom Landecker for interesting discussions on the modelling of aperture
illumination functions.

\end{acknowledgements}

\facilities{VLA, ALMA, MeerKAT, ilifu}

\bibliography{references}

\end{document}